\documentclass[final,number,sort&compress]{elsarticle}

\usepackage{amsmath,amssymb}
\usepackage{dcolumn}
\usepackage{subcaption}
\usepackage[margin=2.5cm]{geometry}
\usepackage{hyperref}

\journal{Int. J. Heat Fluid Flow -- VoR DOI: \href{https://doi.org/10.1016/j.ijheatfluidflow.2021.108889}{10.1016/j.ijheatfluidflow.2021.108889}}

\begin{document}

\begin{frontmatter}

\title{Stochastic modeling of surface scalar-flux fluctuations in turbulent channel flow using one-dimensional turbulence}

\noindent \textcopyright\ 2021. 
Licensed under the Creative Commons CC-BY-NC-ND 4.0 license \\ 
\url{https://creativecommons.org/licenses/by-nc-nd/4.0/}

\author{Marten Klein\corref{cor1}}

\author{Heiko Schmidt}

\address{Lehrstuhl Numerische Str\"omungs- und Gasdynamik, Brandenburgische Technische Universit\"at Cottbus-Senftenberg, Cottbus, Germany}

\author{David O. Lignell}

\address{Chemical Engineering, Brigham Young University, Provo, Utah, U.S.A.}

\cortext[cor1]{Corresponding author.}

\begin{abstract}
Accurate and economical modeling of near-surface transport processes is a standing challenge for various engineering and atmospheric boundary-layer flows.
In this paper, we address this challenge by utilizing an economical stochastic one-dimensional turbulence (ODT) model.
ODT aims to resolve all relevant scales of a turbulent flow for a one-dimensional domain.
Here ODT is applied to turbulent channel flow as stand-alone tool.
The ODT domain is a wall-normal line that is aligned with the mean shear.
The free model parameters are calibrated once for the turbulent velocity boundary layer at a fixed Reynolds number.
After that, we use ODT to investigate the Schmidt ($Sc$), Reynolds ($Re$), and Peclet ($Pe$) number dependence of the scalar boundary-layer structure, turbulent fluctuations, transient surface fluxes, mixing, and transfer to a wall.  
We demonstrate that the model is able to resolve relevant wall-normal transport processes across the turbulent boundary layer and that it captures state-space statistics of the surface scalar-flux fluctuations.
In addition, we show that the predicted mean scalar transfer, which is quantified by the Sherwood ($Sh$) number, self-consistently reproduces established scaling regimes and asymptotic relations with respect to $Sc$, $Re$, and $Pe$.
For high asymptotic $Sc$ and $Re$, ODT results fall between the Dittus--Boelter, $Sh\sim Re^{4/5}\,Sc^{2/5}$, and Colburn, $Sh\sim Re^{4/5}\,Sc^{1/3}$, scalings but they are closer to the former.
For finite $Sc$ and $Re$, the model prediction reproduces the relation proposed by Schwertfirm and Manhart (\emph{Int. J. Heat Fluid Flow}, \textbf{28}, 1204--1214, 2007) that is based on boundary-layer theory and yields a locally steeper effective scaling than any of the established asymptotic relations. 
The model extrapolates the scalar transfer to small asymptotic $Sc\ll Re_\tau^{-1}$ (diffusive limit) with a functional form that has not been previously described.
\end{abstract}


\end{frontmatter}


\section{\label{sec:intro} Introduction}

Numerical modeling of scalar transport in turbulent boundary layers is a standing challenge that is relevant for a wide range of applications from small flames to large atmospheric flows.
Key problems are related to small-scale correlations, scale interactions, and counter-gradient fluxes (e.g.~\cite{Abe_Kawamura_Matsuo:2004,Pirozzoli_etal:2016,Deardorff:1966}).
Due to the latter, all relevant scales of the flow have to be resolved for robust numerical predictions.
Direct numerical simulation (DNS) is the ideal tool, but it is of limited applicability due to the resolution requirements imposed by the Kolmogorov and Batchelor scales (e.g.~\cite{Hasegawa_Kasagi:2009,Schwertfirm_Manhart:2007,Ostilla-Monico_etal:2015}). 
State-of-the-art DNS are unfortunately only feasible for modestly varying Reynolds and Schmidt (or Prandtl) number ranges (e.g.~\cite{Abe_Antonia:2017,Abe_Antonia:2019,Alcantara-Avila_etal:2018,Alcantara-Avila_Hoyas:2021,Alcantara-Avila_etal:2021}). 

Therefore, the development of feasible but physically accurate fluctuation resolving flow and transport models has been identified as a key challenge (e.g.~\cite{Saxton-Fox_McKeon_TSFP:2017,Ebadi_etal:2020,Ranjan_Menon:2021}).  
Here we address the numerical challenge of fluctuation modeling  for robust extrapolation of scaling laws of the scalar transfer by utilizing an economical stochastic one-dimensional turbulence (ODT) model \cite{Kerstein:1999,Kerstein_etal:2001}.
It is worth to point out that small-scale near-wall processes are usually not resolved even in state-of-the-art large-eddy simulation (LES) such that predictive capabilities can be limited (e.g.~\cite{Piomelli:2008}). 
In these LES, the near-wall flow is either modeled with the aid of additional closure equations for the wall-shear stress and heat flux (so-called wall-modeled LES; WMLES), or resolved only in an average sense by performing Reynolds-averaged Navier--Stokes simulations (RANS) for the near-wall region. 
In RANS, only mean effects are represented and these are modeld by prescribed universal wall functions. 
These limitations do not apply to ODT. 
There is no closure and, hence, no closure modeling, even for the smallest scales in the flow close to the wall.

ODT has been validated from a fundamental point of view and applied to multi-physics boundary layers (e.g.~\cite{Kerstein_Wunsch:2006,Medina_etal:2019,Monson_etal:2016,Fragner_Schmidt:2017}).
In such applications, ODT aims to provide full resolution of the whole range of relevant scales with modeling in the physical coordinate.
Based on the capturing of the momentum boundary layer \cite{Kerstein:1999}, it has been assumed that ODT `automatically' captures also scalar transport processes with similar fidelity, with ensuing applications, but this was never systematically investigated.
In this paper, we extend two preliminary studies \cite{Klein_Schmidt_TSFP:2017,Klein_Schmidt_STAB:2021} and test the assumption mentioned. 
We also discuss the limitations of the simple model so that it can be more confidently applied. 
This is specifically important as ODT is used for model development and simulation, for instance, as subgrid model in large-eddy simulation \cite{Schmidt_etal:2003,Gonzalez-Juez_etal:2011,Glawe_etal:2018}, among other uses.

\begin{figure}[tp]
  \centering
  \includegraphics[height=42mm]{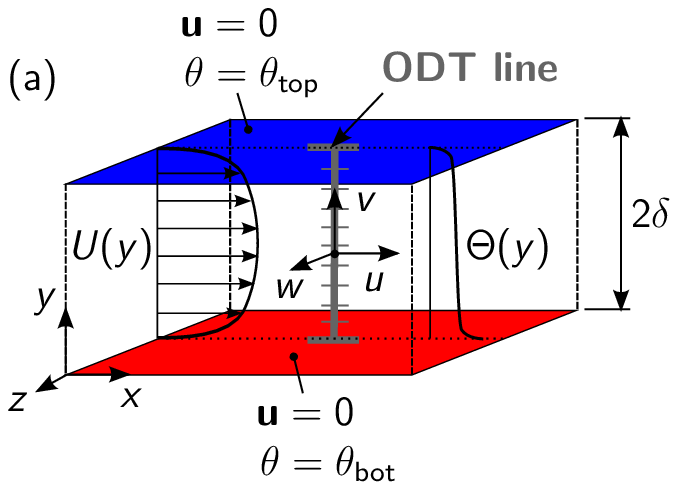}
  \includegraphics[height=42mm]{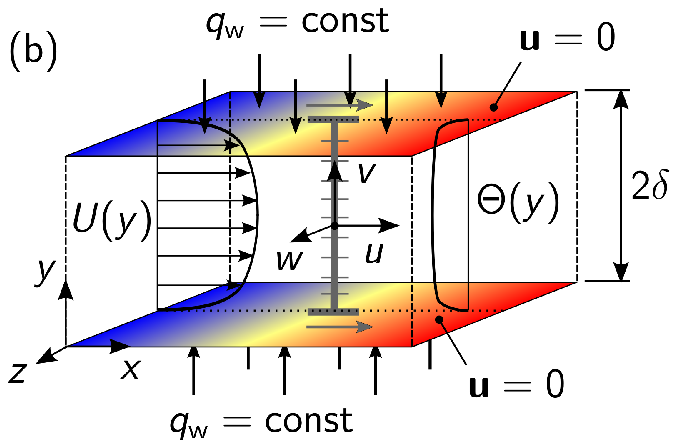}
  \caption{%
    Schematic of the channel flow configuration and the scalar forcing mechanisms investigated.
    ODT simulations are carried out for a lower-order computational domain, the so-called `ODT line', that spans the whole channel of height $h=2\delta$.
    (a)~Constant scalar value (CSV) forcing with fixed wall values $\theta_\text{bot}>\theta_\text{top}$.
    (b)~Constant scalar flux (CSF) forcing with prescribed isoflux wall-boundary condition for which an ODT line nominally moves downstream with the bulk velocity during a simulation.
  }
  \label{fig:config}
\end{figure}

Here we apply the model to canonical channel flow configurations with height $h=2\delta$, as sketched in figure~\ref{fig:config}, and investigate the Schmidt and Reynolds number dependence of the bulk-surface coupling in terms of the scalar transfer to the wall by distinguishing molecular (diffusive) and turbulent (advective) wall-normal transport processes.
ODT facilitates the analysis of fluctuating surface scalar and momentum fluxes and their relation to low-order flow statistics in the turbulent boundary layer.
For the analysis of scalar transport, we focus on a passive scalar that is given by weak temperature variations or a chemical tracer that has negligible effect on the mass, momentum, and energy balances \cite{Monin_Yaglom:1971}.
Most of the results shown below have been obtained for constant scalar value (CSV) wall-boundary conditions (that is, isothermal walls), but we also consider the complementary case of a prescribed constant scalar flux (CSF; that is, a heated channel).

This paper is organized as follows.
The ODT model is formulated in section~\ref{sec:method}.
We then summarize the model application and calibration for channel flow in section~\ref{sec:val}.
In section~\ref{sec:results}, we report and discuss model predictions for passive scalar turbulence statistics in the boundary layer and their relation to the state-space of surface flux fluctuations as well as the mean scalar transfer to a wall.
Section~\ref{sec:conc} concludes the paper, and some additional material is presented in the appendix.

\section{\label{sec:method} Model formulation}

Kertein's ODT model is a stochastic model of turbulent flow that resolves all relevant length and time scales in a single spatial dimension. This results in a high-fidelity model that is computationally efficient in comparison to DNS. This section provides an overview of the ODT model with a focus on the most relevant aspects for the present application. Details of the formulation and its implementation can be found in \cite{Kerstein:1999,Kerstein_etal:2001,Lignell_etal:2013}. For convenience, a brief but complete presentation of the model equations is provided in \ref{sec:odt-appendix}.

\begin{figure}[tp]
  \centering
  \includegraphics[height=52mm]{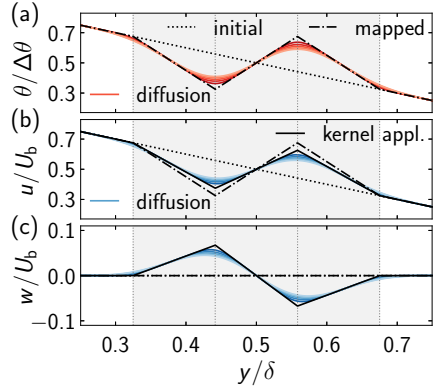}
  \caption{%
    Visualization of the ODT temporal advancement for (a)~the passive scalar, (b)~the streamwise, and (c)~the spanwise velocity, respectively.
    A discrete eddy event exhibits instantaneous triplet mapping (dash-dot) and kernel application (solid) across a finite-size interval (shaded). It is followed by deterministic diffusion (color). 
    Initial scalar, $\theta(y,t_0)$, and streamwise velocity, $u(y,t_0)$, profiles (dotted) are here linear and normalized by the bulk scalar difference, $\Delta\theta$, and bulk velocity, $U_\text{b}$, respectively.
    The spanwise, $w(y,t_0)$, and wall-normal, $v(y,t_0)$ (not shown), velocity profiles are initially zero, but receive  the same amount of kinetic energy by the kernel application.
    The three segments of the triplet map are separated by dotted vertical lines.
  }
  \label{fig:seq_tm}
\end{figure}

The ODT domain, can be thought of as a line-of-sight through a turbulent flow on which momentum and other scalar fields evolve. 
The model solves one-dimensional, unsteady, deterministic PDEs for momentum components and other scalar fields. 
These equations, presented below and referred to as the diffusion equations, include molecular transport and source terms. 
The one-dimensional parabolic PDEs are suitable for solution of a wide range of boundary-layer-like flows, such as jets, mixing layers, and channel flows, which are assumed to be statistically one-dimensional. 
Turbulent advection is inherently three-dimensional and cannot be directly represented with these equations. Instead, the effect of turbulent advective transport on the scalar profiles is modeled through stochastic eddy events. 
These eddy events are instantaneous and implemented with a domain re-mapping process called a triplet map, discussed below and illustrated in figure~\ref{fig:seq_tm}. 
The eddy events are characterized by three random variables: the eddy size $l$, location $y_0$, and time-of-occurrence $t_\text{e}$. 
These eddy events are sampled from an eddy rate probability density function (PDF) that depends directly on the dynamically-evolving momentum fields, and so varies with time and position. 
Given an initial condition, the next eddy event (size, location, and time) is sampled and implemented (meaning the domain profiles are remapped), and the diffusion equations are solved up to the eddy event time. 
The process is then repeated to the end time. 
Multiple flow realizations are computed in order to gather statistics for data analysis. 

The triplet map is defined mathematically in equation~(\ref{eq:triplet}), and qualitatively as follows. 
At location $y_0$ and eddy size $l$, the triplet map is implemented by making three copies of all domain profiles, compressing them spatially by a factor of three, arranging the three copies sequentially, and then spatially inverting the middle copy. 
This is shown in figure~\ref{fig:seq_tm}. 
The triplet map is measure-preserving (conserves all quantities and their moments), continuous, increases profile gradient magnitudes, and is local in the sense that a scale reduction occurs by a constant factor for a given eddy size. 
It is nonlocal in the sense that fluid elements are effectively permuted along the one-dimensional computational domain in order to represent the effects of turbulent stirring motions. 
These properties are important for a physically-consistent model of cascade phenomenology.

Figure~\ref{fig:seq_tm} illustrates an eddy event and subsequent diffusive advancement for initially linear scalar and velocity component profiles.
The scalar and velocity profiles are triplet mapped, as noted above, but momentum components are further adjusted by adding a scaled kernel function to represent the effects of three-dimensional pressure-velocity coupling and return-to-isotropy effects \cite{Kerstein_etal:2001} (see \ref{sec:odt-appendix}).
These pressure fluctuations manifest themselves by a redistribution of inter-component kinetic energy, which is mimicked by the kernel application as shown in figures~\ref{fig:seq}(b,c).
Following an eddy event, the scalar and velocity profiles are evolved by the diffusive advancement equations, shown in the blue and red curves in the figure (which do not include the pressure gradient source term for streamwise velocity here). 

The eddy rate $\lambda(l,y_0;t)$ is defined in equation~\ref{eq:eddyRate} and is inversely proportional to an eddy timescale. 
The proportionality constant $C$ is a calibrated model parameter that scales the overall rate of eddies and the flow evolution.
The eddy timescale is written in terms of a measure of the local turbulent kinetic energy $E$ on dimensional grounds, $\tau\sim\sqrt{E/l^2}$. 
To suppress unphysically small eddies, the energy is decreased by a viscous penalty $E_\text{vp}=Z\nu^2/l^2$, where $Z$ is another calibrated parameter. 
For details, see equation~(\ref{eq:eddyTau}) and the corresponding discussion.  

The ODT governing equations for momentum and a passive scalar in turbulent channel flow are given by 
\begin{subequations}
 \renewcommand{\theequation}{\theparentequation \textit{\alph{equation}}}
 \label{eq:gov}
 \begin{align}
  \frac{\partial \boldsymbol{u}}{\partial t} + \sum_{t_\text{e}} \mathcal{E}_{\boldsymbol{u}}(\boldsymbol{u})\,\tilde{\delta}(t-t_\text{e}) &= \nu \frac{\partial^2 \boldsymbol{u}}{\partial y^2} - \frac{1}{\rho} \frac{\text{d}\bar{p}}{\text{d}x}\,\boldsymbol{e}_x \;,
  \\
  \frac{\partial \theta}{\partial t} + \sum_{t_\text{e}} \mathcal{E}_{\theta}(\boldsymbol{u})\,\tilde{\delta}(t-t_\text{e}) &= \Gamma \frac{\partial^2 \theta}{\partial y^2} + s_\theta \;,
 \end{align}
\end{subequations}
where $\boldsymbol{u}=(u,v,w)^\text{T}$ denotes the velocity vector and its Cartesian components, $\theta$ the scalar, $t$ the time, $x$ the stream-wise and $y$ the vertical coordinate, $\rho$ and $\nu$ the fluid's density and kinematic viscosity, $\Gamma$ the scalar diffusivity, $\text{d}\bar{p}/\text{d}x$ the prescribed mean pressure gradient, $\boldsymbol{e}_x$ the unit vector in stream-wise direction, and $s_\theta$ is a scalar source term. 
The second term on the left-hand side of the equations is a symbolic representation of the ensemble effects of three-dimensional turbulent advection that is modeled by the above-mentioned stochastic sequence of discrete eddy events $\mathcal{E}$. The Dirac $\tilde{\delta}$ functions indicate the instantaneous nature of the eddy events. Hence, these equations, symbolically, represent the two concurrent ODT processes of the stochastic eddy sampling and solution of the diffusion equations between the instantaneous eddy events. The diffusion equations are simply equations~(\ref{eq:gov}\textit{a,b}) without the $\mathcal{E}$ terms.
The subscripted notation $\mathcal{E}_{\boldsymbol{u}}$ and $\mathcal{E}_\theta$ is adopted to show that the model distinguishes velocity (momentum) and scalar transport, which is discussed in more detail below.

A suitable dimensionless form of the governing equations~(\ref{eq:gov}\textit{a,b}) may be obtained with the aid of the mean wall-stress balance and the friction scalar property (temperature or concentration),
\refstepcounter{equation}
$$
  u_\tau^2 = \nu\left|\frac{\text{d}U}{\text{d}y}\right|_\text{w} = \frac{\delta}{\rho}\left|\frac{\text{d}\bar{p}}{\text{d}x}\right| \;,
  \qquad
  \theta_\tau = \frac{\Gamma}{u_\tau}\left|\frac{\text{d}\Theta}{\text{d}y}\right|_\text{w} \;,
  \eqno{(\theequation{\mathit{a},\mathit{b}})}
  \label{eq:nondim}
$$
where the upper-case symbols $U(y)=\overline{u(y,t)}$ and $\Theta(y)=\overline{\theta(y,t)}$, as well as the bar, $\overline{(\cdot)}$, denote conventional temporal (Reynolds) averages, $u_\tau$ and $\theta_\tau$ are the friction velocity and scalar property, respectively, and the subscript `$\text{w}$' indicates evaluation at the wall. 
Scaling equations~(\ref{eq:gov}\textit{a,b}) with the friction variables yields 
\begin{subequations}
 \renewcommand{\theequation}{\theparentequation \textit{\alph{equation}}}
 \label{eq:gov2}
 \begin{align}
  \frac{\partial \boldsymbol{u}^+}{\partial t^+} + \sum_{t^+_\text{e}} \mathcal{E}_{\boldsymbol{u}}^+(\boldsymbol{u}^+)\,\tilde{\delta}(t^+-t^+_\text{e}) &= \frac{1}{Re_\tau} \frac{\partial^2 \boldsymbol{u}^+}{\partial y^{+\,2}} + \boldsymbol{e}_x \;,
  \\
  \frac{\partial \theta^+}{\partial t^+} + \sum_{t^+_\text{e}} \mathcal{E}^+_\theta(\boldsymbol{u}^+)\,\tilde{\delta}(t^+-t^+_\text{e}) &= \frac{1}{Sc\,Re_\tau} \frac{\partial^2 \theta^+}{\partial y^{+\,2}} + s_\theta^+ \;,
 \end{align}
\end{subequations} 
where $y^+=y u_\tau/\nu$, $t^+=t u_\tau/\delta$, $u^+=u/u_\tau$, and $\theta^+=(\theta-\theta_\text{w})/\theta_\tau$ denote variables in inner scaling and $\theta_\text{w}$ is a reference (wall) value of the scalar property.
Nondimensional similarity solutions are obtained with dependence on the friction Reynolds, $Re_\tau=u_\tau\delta/\nu$, and Schmidt (Prandtl), $Sc=\nu/\Gamma$, number. 

No-slip wall-boundary conditions are prescribed for the velocity vector, whereas fixed-value and isoflux boundary conditions are prescribed for the scalar, respectively.
The boundary conditions for the scalar define the case set-up, which is addressed in more detail in \ref{sec:bc}.

\section{\label{sec:val} Model application to turbulent channel flow}

In this section we address the general model application to channel flow and some fundamental properties of the stochastic model for the velocity boundary layer.
We do this in order to calibrate the free model parameters and in order to quantify the model's ability to capture turbulent fluctuations.

ODT simulations of turbulent channel flows are conducted as follows. 
Equations~(\ref{eq:gov}\textit{a,b}) are numerically integrated which yields a time sequence of synthetic but statistically representative flow profiles.
Conventional statistics are performed on these profiles and gathered on a predefined post-processing grid for the fully-adaptive solver \cite{Lignell_etal:2013} used.
While cumulative statistics are straightforward, the computation of the ODT-resolved turbulent fluxes (cross-correlations) of fluctuating quantities is conditional on the eddy events and detailed in \ref{sec:cross-corr}. 

We have performed `pre-simulations' for isothermal flow conditions at various $Re_\tau$ in order to calibrate the free model parameters $C$, $Z$, and $\alpha$.
The individual influence of these parameters on the velocity boundary-layer statistics is discussed, for example, in \cite{Rakhi_etal:2019,Schmidt_etal:2003} and not repeated here.
We select $C=6$, $Z=300$, and $\alpha=1/6$ as reasonable model parameter values for the presently investigated $Re_\tau$ range. 
The calibrated model parameters are kept fixed from here on.
We comment on relevant aspects of the model parameter selection and the representation of the viscous boundary layer in \ref{sec:cal}.

\begin{figure}[tp]
  \centering
  \includegraphics[height=48mm]{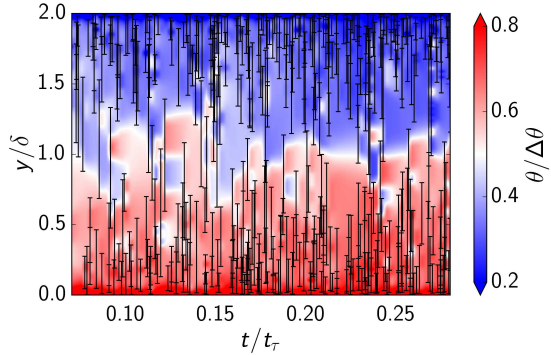}
  \caption{%
    Space-time diagram of ODT eddy events (black lines) together with a $Sc=3$ passive scalar (color shading) in a channel flow with $Re_\tau=180$.
    The scalar is prescribed by fixed wall values (CSV forcing) and perturbed on multiple scales by the stochastically sampled and instantaneously implemented eddy events in between which only molecular diffusion occurs.}
  \label{fig:seq}
\end{figure}

At this point we briefly describe how the discrete stochastic (turbulent advective) and continuous deterministic (molecular diffusive) transport processes interact. 
Figure~\ref{fig:seq} shows a space-time diagram of an ODT realization of a turbulent channel flow with $Re_\tau=180$ and $Sc=3$.
The scalar is prescribed by fixed wall values (CSV) that differ by $\Delta\theta=\theta_\text{bot}-\theta_\text{top}>0$.
The scalar field fluctuates notably in the bulk due to application of the triplet maps. 
The mapping intervals form a random sequence and are given as black vertical lines.
Intense turbulent mixing occurs where lines of different size cluster. 
Note that eddy events are mostly `anchored' in the near-surface region for turbulent channel flow, which is the ODT representation of wall-attached eddies (e.g.~\cite{Townsend:1976,Marusic_Monty:2019}).
Note that it is a dynamical feature of the ODT formulation that captures these physical properties qualitatively, which is discussed next.

ODT simulations need to sufficiently sample the state space. 
We therefore ran long-time simulations with $t_\text{a}u_\tau/\nu\gg1$, where $t_\text{a}$ is the simulated averaging time interval per realization and $\delta/u_\tau$ a statistical eddy turnover time.
If necessary, $t_\text{a}$ can be significantly reduced by adopting large $M$ ensembles of flow realizations.
The numerical solver used is fully adaptive such that the stochastic ODT simulations performed were permissively resolving a boundary-layer diffusive length scale, that is, the Kolmogorov or Batchelor scale, of the wall-bounded flow for $Sc\leqslant1$ and $Sc>1$, respectively.
The minimum grid cell size is $\Delta y_\text{min}$ and is assured to be smaller than the estimated Kolmogorov and Batchelor scales.
Table~\ref{tab:config} summarizes relevant details and bulk quantities for a representative set of ODT simulations.

\begin{table}[thp]
  \centering
  \caption{%
    Details of ODT simulation parameters and bulk quantities for representative cases with two different scalar wall-boundary conditions ($\theta$-BC).
    $Re_\tau$ and $Re$ denote the friction and bulk Reynolds number, respectively; $Sc$ the Schmidt number; $K^+=\theta_\tau/\Delta\theta=Sh/(2\,Sc\,Re_\tau)$ the scalar transfer coefficient, where $\Delta\theta=|\overline{(\theta_\text{b}-\theta_\text{w})}|$ is the mean scalar bulk-wall difference, $\theta_\tau$ the scalar wall unit, and $Sh$ the Sherwood (Nusselt) number that is detailed in section~\ref{sec:Sh}; $\bar{N}_y$ the average number of cells in the adaptive grid; $\Delta y_\text{min}^+$ the normalized minimum allowed cell size; and $t_\text{s}u_\tau/\delta$ the simulated eddy turnover times per realization. 
    Core hours (CPU-h) spent per realization are measured on Intel\textsuperscript{\textregistered} Xeon\textsuperscript{\textregistered} E5-2630 ($2.40\,\text{GHz}$) processors.
  }
  \begin{tabular}{D{.}{.}{+1} r c l c r l r D{.}{.}{-1}} 
    \hline
    \\
    \multicolumn{1}{c}{$Re_\tau$} & \multicolumn{1}{c}{$Re$} & \multicolumn{1}{c}{$Sc$} & 
    \multicolumn{1}{c}{$K^+$} & \multicolumn{1}{c}{$\theta$-BC} & 
    \multicolumn{1}{c}{$\bar{N}_y$} & \multicolumn{1}{c}{$\Delta y_\text{min}^+$} & 
    \multicolumn{1}{c}{$t_\text{s}u_\tau/\delta$} & \multicolumn{1}{c}{CPU-h} \\
    \\
     180.4 &      2660 &    0.025 & 0.275    & CSV &  166 & 0.2   & 15{,}000 &   5.5 \\
     180.4 &      2663 &    0.71  & 0.0475   & CSV &  163 & 0.2   & 15{,}000 &   5.5 \\
     180.2 &      2665 &   10     & 0.0150   & CSV &  164 & 0.2   & 15{,}000 &   7.5 \\
     180.0 &      2716 &  200     & 0.00251  & CSV &  170 & 0.036 &     1500 &  20   \\
     180.0 &      2723 & 1000     & 0.000893 & CSV &  174 & 0.018 &     1500 &  68   \\
    \\%
     591.5 &  10{,}806 &    0.025 & 0.130    & CSV &  207 & 0.12  & 4500 &   5.5 \\
     591.5 &  10{,}809 &    0.71  & 0.0426   & CSV &  203 & 0.12  & 4500 &   5.5 \\
     590.3 &  10{,}890 &   10     & 0.0147   & CSV &  198 & 0.12  & 4500 &   7.5 \\
     590.0 &  11{,}216 &  200     & 0.00255  & CSV &  207 & 0.024 &  450 &  25   \\
     589.4 &  11{,}360 & 1000     & 0.000907 & CSV &  218 & 0.012 &  450 &  97   \\
    \\%
     994.2 &  19{,}126 &    0.025 & 0.164    & CSF &  223 & 0.36  & 2000 &  94   \\
     994.5 &  19{,}133 &    0.71  & 0.0523   & CSF &  225 & 0.36  & 2000 &  94   \\
    \\%
    2007   &  43{,}552 &    0.025 & 0.0827   & CSV &  273 & 0.4   & 1200 &  93   \\
    2008   &  43{,}607 &    0.71  & 0.0381   & CSV &  285 & 0.4   & 1200 &  94   \\
    2007   &  43{,}599 &   10     & 0.0139   & CSV &  269 & 0.4   & 1200 &  56   \\
    1993   &  44{,}563 &  200     & 0.00243  & CSV &  871 & 0.08  &  300 &  33   \\
    2017   &  43{,}237 & 1000     & 0.000862 & CSV & 1195 & 0.08  &  300 &  38   \\
    \\%
    5222   & 125{,}750 &    0.025 & 0.0655   & CSV &  526 & 0.21  &  480 &  31   \\
    5223   & 125{,}874 &    0.71  & 0.0351   & CSV &  551 & 0.21  &  480 &  31   \\
    5213   & 126{,}803 &   10     & 0.0139   & CSV & 2737 & 0.1   &  480 &  19   \\
    5271   & 125{,}370 &  200     & 0.00252  & CSV & 3256 & 0.1   &  240 &  64   \\
    5269   & 125{,}411 & 1000     & 0.000904 & CSV & 7144 & 0.1   &  120 &  33   \\
    \\%
    \hline
  \end{tabular}    
  \label{tab:config}
\end{table}

\section{\label{sec:results} Results and discussion}

In the following, we investigate the scalar boundary layer, fluctuations and wall-normal fluxes using conventional statistics.
We then turn to the detailed statistics of the transient surface scalar and momentum fluxes which govern the fluid-surface coupling but are commonly not resolved in filter-based approaches such as LES, WMLES, and RANS.
Next, the modeled scalar fluctuation variance and mixing properties are investigated for wall-bounded turbulence. 
Finally, we address ODT's capabilities for extrapolation by investigating the scaling regimes of the mean surface scalar transfer.
This is done with the calibrated model set-up and primarily for fixed-value (CSV) forcing of the scalar.
Isoflux (CSF) forcing is used to demonstrate boundary layer similarity and for direct comparisons with available reference data.

\subsection{\label{sec:tmean} Scalar boundary layer}

Recently, evidence from DNS was presented that the dissimilarity of scalar and momentum transport manifests itself also in the boundary layer structure \cite{Hasegawa_Kasagi:2009,Abe_Antonia:2017} which has been long known from measurements (e.g.~\cite{Shaw_Hanratty:1977,Kader:1981}).
DNS resolve the intricate dynamics in the turbulent boundary layer and have thus been the tool of choice (e.g.~\cite{Kasagi:1992,Owinoh_etal:2005,Pirozzoli_etal:2016}). 
The question addressed in the following, therefore, is to what extend the lower-order ODT model is able to capture the scalar boundary layer structure?

\begin{figure}[tp]
  \centering
  \includegraphics[height=52mm]{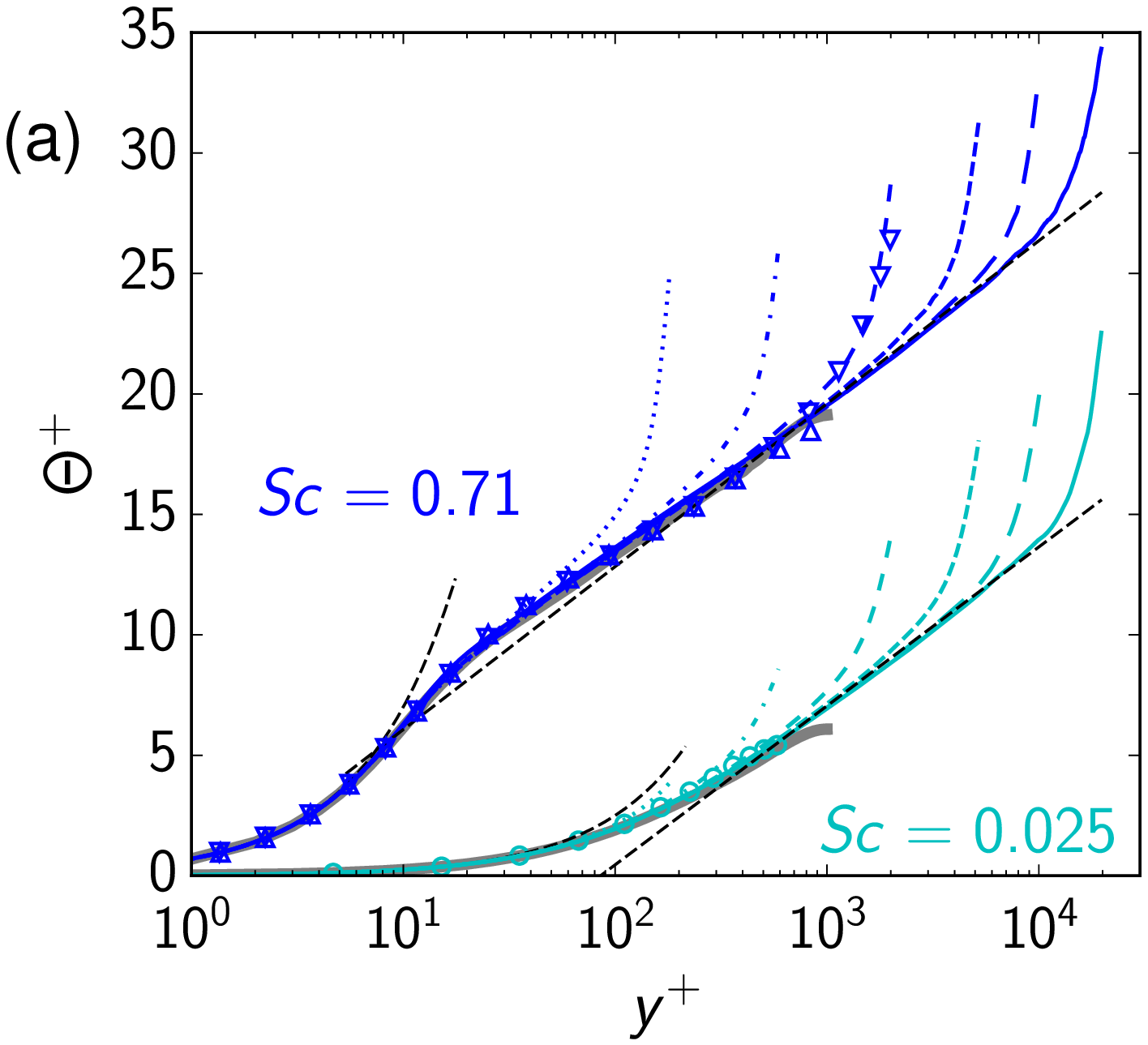}
  \includegraphics[height=52mm]{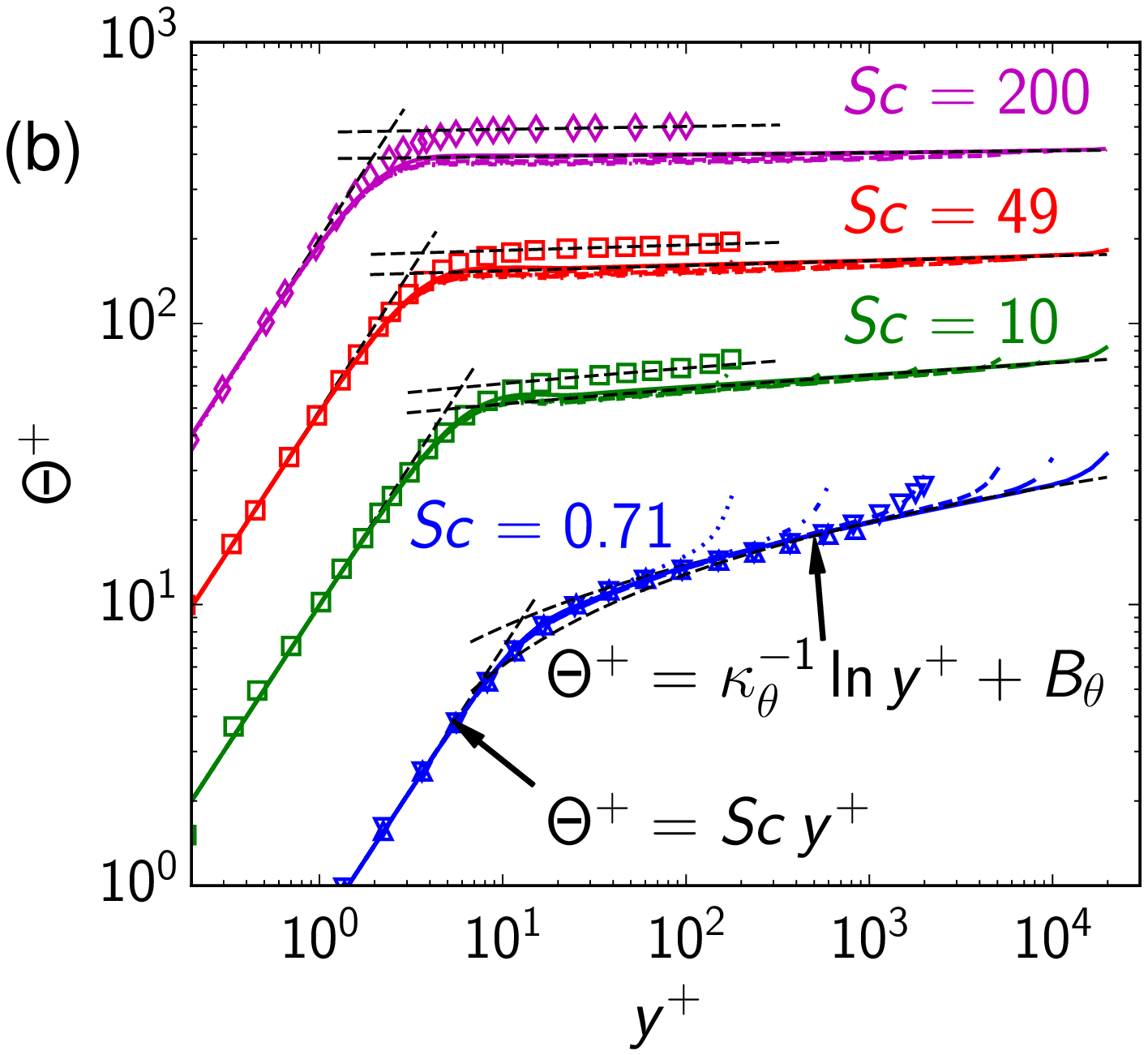}
  \\
  \includegraphics[height=52mm]{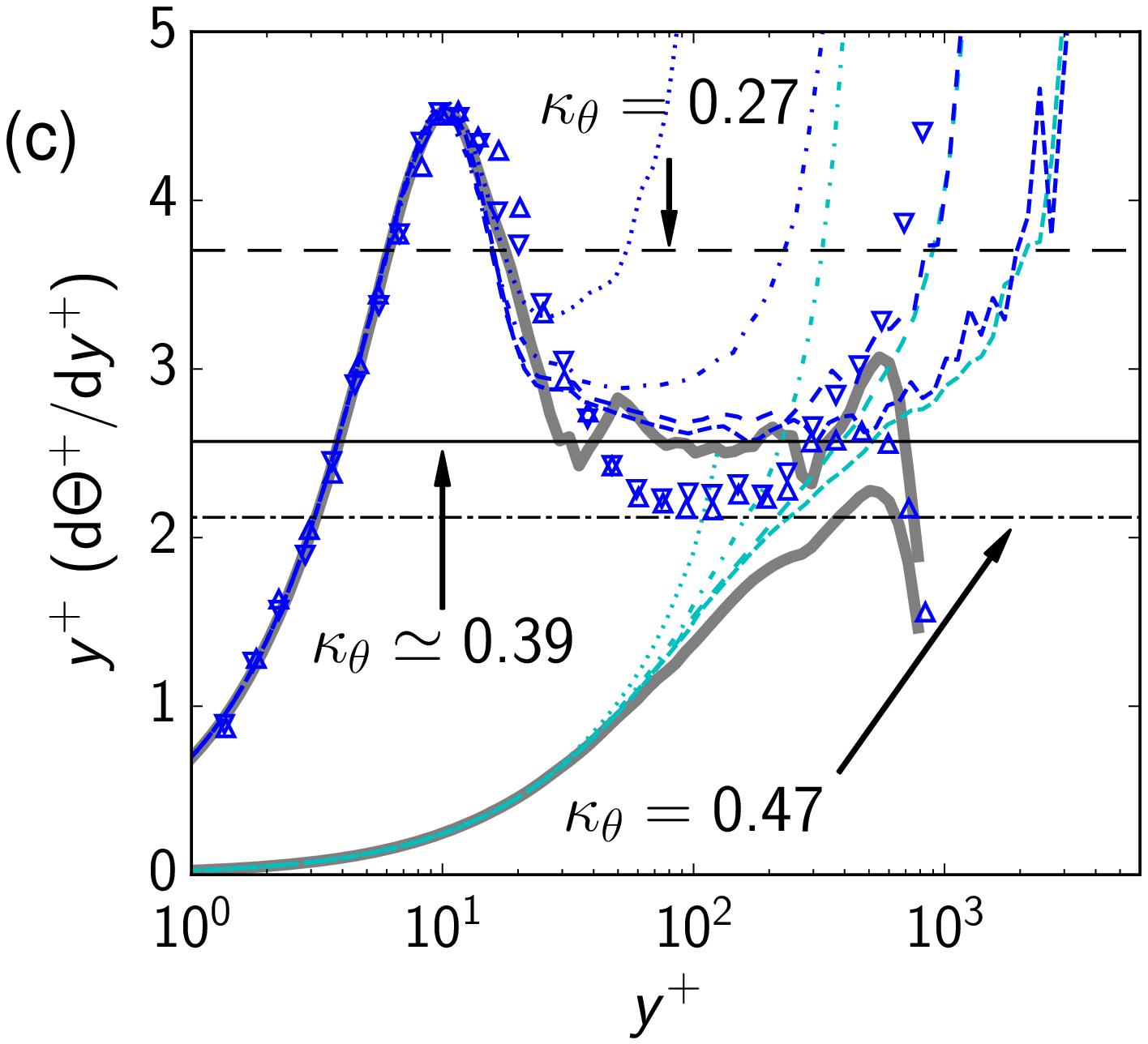}
  \includegraphics[height=52mm]{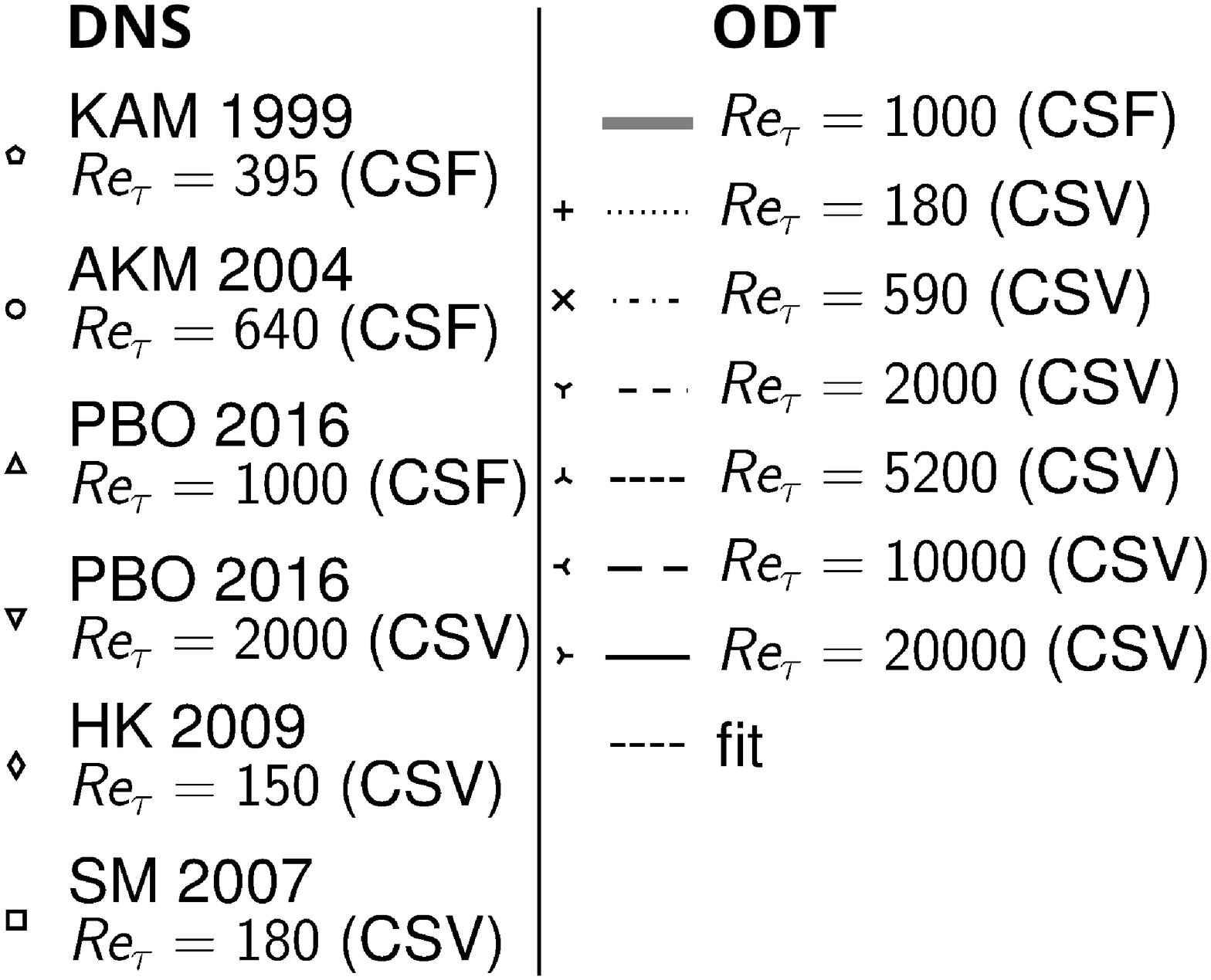}
  \caption{%
    Mean profiles of the normalized scalar, $\Theta^+$, for various $Sc$ and $Re_\tau$;
    (a)~low $Sc\leqslant0.71$ regime in semi-logarithmic, and
    (b)~high $Sc\geqslant0.71$ regime in double-logarithmic scale.
    (c)~Indicator function, $y^+\,\left(\text{d}\Theta^+/\text{d}y^+\right)$, corresponding to panel~(a) but limited to $Re_\tau\leqslant5200$ for which various reference data are available.
    Open symbols denote reference DNS from \cite{Kawamura_Abe_Matsuo:1999,Abe_Kawamura_Matsuo:2004,Pirozzoli_etal:2016,Hasegawa_Kasagi:2009,Schwertfirm_Manhart:2007}.
    Line styles (as well as cross-like symbols in figures below) distinguish ODT results for different $Re_\tau$.
    Colors distinguish different $Sc$.
    Fits of approximately linear and logarithmic regions are given by solid and dashed black lines.
  }
  \label{fig:tmean}
\end{figure}

Wall-normal profiles of the normalized scalar mean field, $\Theta^+=(\bar{\theta}-\theta_\text{w})/\theta_\tau$, are shown for various $Re_\tau$ in figure~\ref{fig:tmean}(a) and \ref{fig:tmean}(b) for $Sc\leqslant0.71$ and $Sc\geqslant0.71$, respectively, where $y^+$ serves as boundary-layer coordinate.
In figure~\ref{fig:tmean}(a), a linear sublayer can be discerned for $y^+<10$ where all data collapse on the diffusive reference curve $\Theta^+(y^+)=Sc\,y^+$ (thin dashed). 
The sublayer extends up to $y^+\approx200$ for the more diffusive cases with $Sc=0.025$.
Hence, a log layer can only be discerned for high enough $Re_\tau$, that is, for friction Peclet number $Pe_\tau=Sc\,Re_\tau\geqslant O(100)$ such that the scalar boundary layer extends beyond $y^+\simeq O(100)$.
The log region is indicated by empirical fits (oblique black dashed lines) to the ODT results with $Re_\tau=20{,}000$ for the range $100<y^+<0.2\,Re_\tau$.
The functional dependence is given by (e.g.~\cite{Kader:1981})
\begin{equation}
  \Theta^+(y^+)=\kappa_\theta^{-1}\,\ln y^+ + B_\theta \;,
  \label{eq:sca-log}
\end{equation}
where $\kappa_\theta$ is the von~K\'arm\'an `constant' and $B_\theta$ the additive `constant' for the scalar.
Both coefficients may retain a dependence on $Sc$ and $Re_\tau$, which is discussed below in more detail.
The boundary layer structure obtained with ODT exhibits a log region for high $Re_\tau$ in agreement with available reference experiments \cite{Kader:1981} and DNS \cite{Pirozzoli_etal:2016,Abe_Kawamura_Matsuo:2004,Abe_Antonia:2019}.
In these DNS, the scalar is either prescribed by CSF ($\Theta^+$ bending downward for large $y^+$) or CSV ($\Theta^+$ bending upward) such that the functional dependence $\Theta^+(y^+)$ takes a different form in the outer layer depending on the scalar forcing \cite{Pirozzoli_etal:2016}, which is fully captured by ODT.

For cases with $Sc\geqslant10$ shown in figure~\ref{fig:tmean}(b), the ODT boundary-layer structure is in reasonable agreement with available low $Re_\tau$ reference DNS \cite{Hasegawa_Kasagi:2009,Schwertfirm_Manhart:2007}.
Only CSV forcing is considered here which is expected to have negligible effect on the inner layer structure for high Reynolds \cite{Pirozzoli_etal:2016,Alcantara-Avila_Hoyas:2021} or, rather, Peclet \cite{Abe_Antonia:2019} numbers.
The scalar is concentrated in the vicinity of the wall, that is, in the diffusive surface layer with $\Theta^+(y^+)=Sc\,y^+$ as indicated.
A log region can be discerned that appears slightly curved due to double-logarithmic axes.
However, the log region runs almost flat for high $Sc$. 
The prominent outer layer `spike', which can be seen at large $y^+>Re_\tau/2$ for $Sc=0.71$ and $0.025$ investigated, disappears for high asymptotic $Sc\gg1$.
For fixed $Sc>1$, the ODT simulated scalar boundary layer profiles collapse for all $Re_\tau$ investigated demonstrating that inner scaling holds up to high asymptotic control parameter values.
There is no indication of a asymptotic $Re_\tau$ dependence of the scalar mean field.
Note also that the model systematically, but proportionally with $Sc$, underestimates $\Theta^+$ for $y^+>10$ throughout the log and outer layers.
This is due to an overestimation of $\theta_\tau$ (compare with table~\ref{tab:config}) that, for fixed scalar wall values (CSV forcing), manifests itself by a systematically lower additive constant $B_\theta$ as shown previously for small Reynolds numbers \cite{Klein_Schmidt_TSFP:2017}.
We discuss below and, additionally, in sections~\ref{sec:jpdf} and \ref{sec:Sh} how this is related to the dissimilarity of the scalar and momentum transfer to the wall that is only partially captured by the model.

Figure~\ref{fig:tmean}(c) shows the indicator function, $y^+\,\left(\text{d}\Theta^+/\text{d}y^+\right)$, that signalizes presence of a log region when
\begin{equation}
  y^+ \, \dfrac{\text{d}\Theta^+}{\text{d}y^+} = \dfrac{1}{\kappa_\theta} = \text{const.}
  \label{eq:indicator}
\end{equation}
Here we limit our attention to $Sc=0.71$ and $0.025$ for $Re_\tau\leqslant5200$ in order to address the relevant aspects of the ODT flow physics representation of the scalar transfer.
For $Sc=0.025$, no plateau can be discerned for the ODT results shown since $Re_\tau=5200$ ($Pe_\tau=130$) is too small to yield a fully turbulent scalar boundary layer.
This is different for $Sc=0.71$, where ODT predicts a plateau at around $\kappa_\theta\simeq0.39$ for $Re_\tau=1000$ ($Pe_\tau=710$) in the case of CSF, but at significantly higher $Re_\tau=5200$ ($Pe_\tau=3700$) in the case of CSV.
Reference experiments from \cite{Kader:1981} yield a universal plateau at $\kappa_\theta\simeq0.47$.  
Recent reference DNS \cite{Abe_Antonia:2019,Alcantara-Avila_etal:2018,Alcantara-Avila_etal:2021} yield a value of $\kappa_\theta\simeq0.43\pm0.02$, which is midway between experiments and present ODT results although for mixed boundary conditions.
These DNS hint at a $Re_\tau$ dependence that is qualitatively reproduced by ODT.
DNS \cite{Pirozzoli_etal:2016,Abe_Antonia:2017,Abe_Antonia:2019} for CSF and CSV suggest the high $Re_\tau>5000$ asymptotic value of $\kappa_\theta\simeq0.46$, which is close to reference experiments \cite{Kader:1981}. 
ODT results for $\kappa_\theta$ suggest a weak but reasonable overestimation of the interaction of the outer and inner layer by a larger degree of similarity between the scalar and momentum transport across the log region.
This aspect is further discussed below.

\begin{figure}[tp]
  \centering
  \includegraphics[height=52mm]{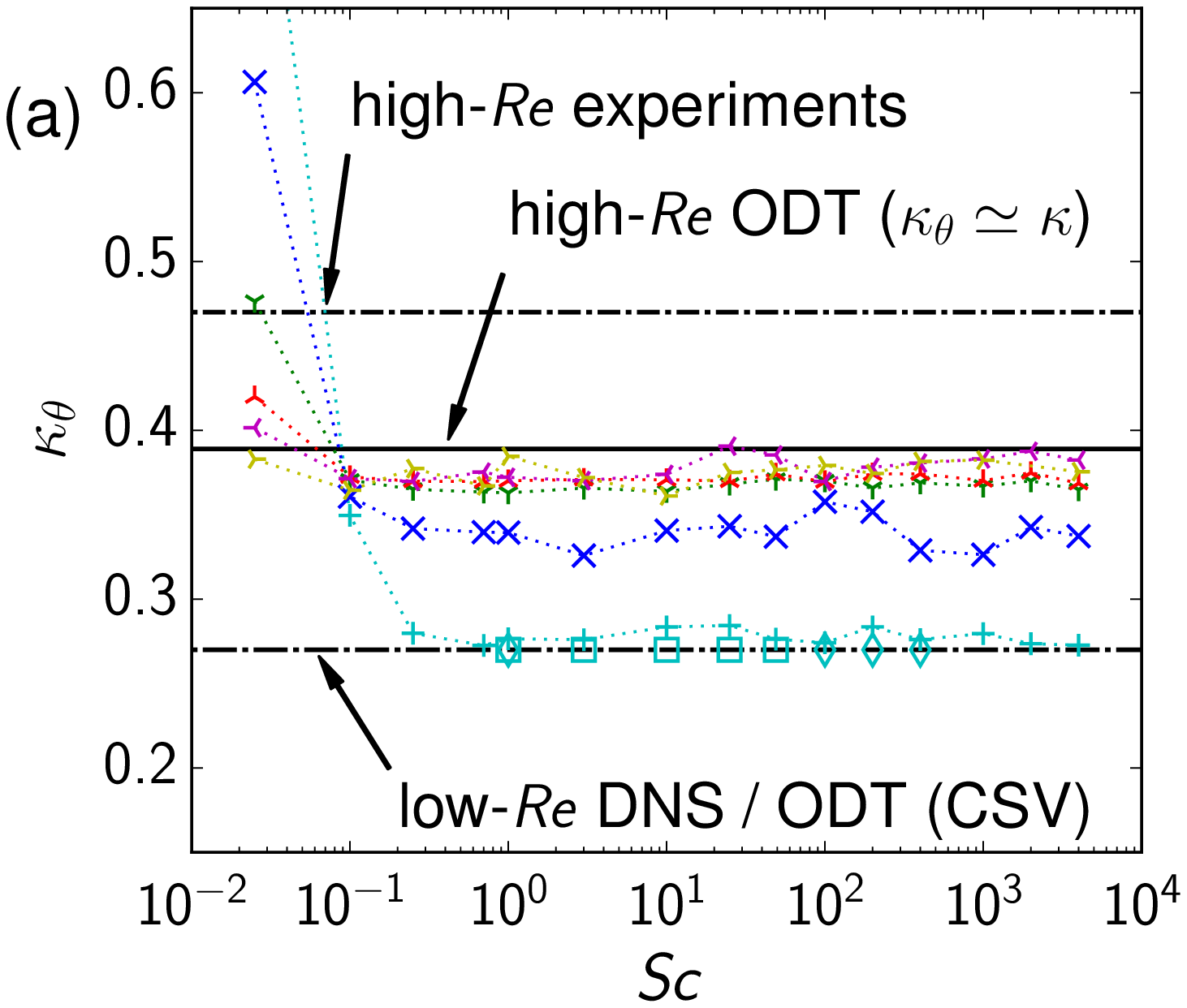}
  \includegraphics[height=52mm]{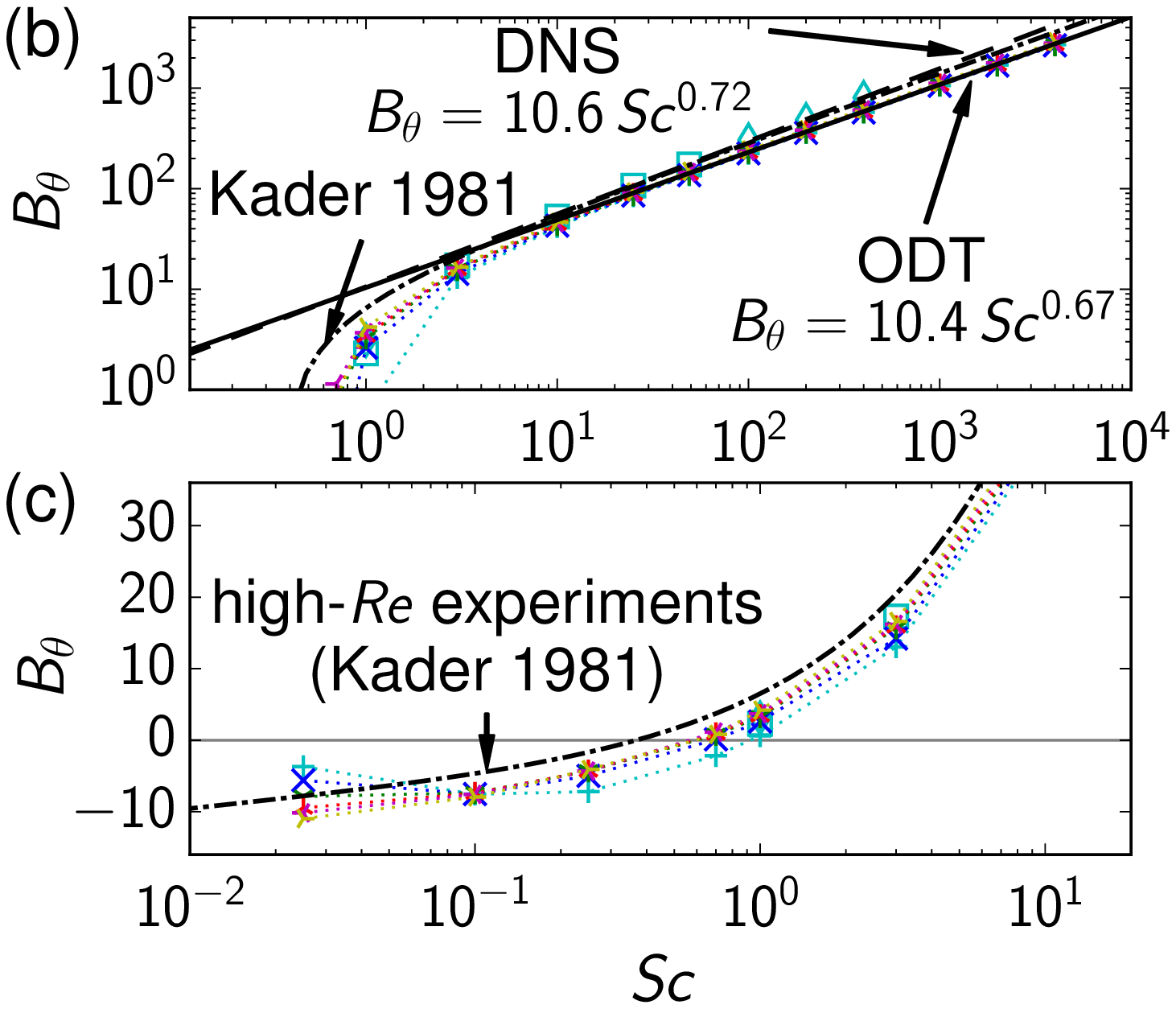}
  \caption{%
    Log layer fit coefficients for various $Re_\tau$ and $Sc$ for CSV forcing.
    (a)~von-K\'{a}rm\'{a}n constant, $\kappa_\theta$, and  
    (b,\,c) additive constant, $B_\theta$, for high and low $Sc$, respectively.
    Open and cross-like symbols distinguish DNS and ODT as indicated in figure~\ref{fig:tmean}.
    Colors are used in addition to visually group the data by $Re_\tau$.
    Empirical relations due to equations~(\ref{eq:BT-Kader}) and (\ref{eq:BT}) are given by broken and solid black lines.
  }
  \label{fig:tsca-fitCoef}
\end{figure}

Fit coefficients $\kappa_\theta$ and $B_\theta$ according to equation~(\ref{eq:sca-log}) are collected in figure~\ref{fig:tsca-fitCoef} for all ODT simulated CSV cases. 
Figure~\ref{fig:tsca-fitCoef}(a) shows increasing $\kappa_\theta$ for decreasing $Sc\ll1$.
This is due to an increasing overlap of the diffusive inner and outer layers due to which the log layer ceases to exist (see figure~\ref{fig:tmean}(c)).
Interestingly, $\kappa_\theta$ retains a notable $Re_\tau$ dependence for high asymptotic $Sc$.
For low $Re_\tau\simeq180$ investigated, $\kappa_\theta=0.27\pm0.02$ for ODT and DNS \cite{Schwertfirm_Manhart:2007,Hasegawa_Kasagi:2009} that were analyzed in the same way.
Note that the fitting range for an assumed \emph{effective} log region was reduced to $40<y^+<0.5\,Re_\tau$ for all DNS and ODT cases with $Re_\tau\leqslant180$.
For high $Re_\tau\geqslant5200$ investigated, ODT predicts $\kappa_\theta=0.37\pm0.02$, but it seems as if $\kappa_\theta$ approaches the asymptotic value $\kappa\simeq0.39$ \cite{Marusic_etal:2010} of the momentum boundary layer.
This is different in reference experiments \cite{Kader:1981} ($\kappa_\theta\simeq0.47$) and DNS \cite{Pirozzoli_etal:2016,Abe_Antonia:2019} ($\kappa_\theta\simeq0.46$) as explained above.

Figures~\ref{fig:tsca-fitCoef}(b,\,c) show the $Sc$ dependence of the additive constant $B_\theta$ for various $Re_\tau$.
Present ODT results suggest that $B_\theta$ does not notably depend on $Re_\tau$ which agrees with the literature. 
An empirical relation for $B_\theta$ is given in \cite{Kader:1981} as 
\begin{align}
  B_\theta(Sc) = (3.85\,Sc^{1/3} - 1.3)^2 + 2.12\, \ln(Sc) 
  \label{eq:BT-Kader}
  \\    
  \quad\text{for}\quad Sc \in\left[6\times10^{-3},4\times10^4\right],
  \notag
\end{align}
where the factor $2.12=\kappa_\theta^{-1}$ is the inverse of the asymptotic reference value $\kappa_\theta\simeq0.47$.
The corresponding ODT prediction is shown in figures~\ref{fig:tsca-fitCoef}(b,\,c) and exhibits satisfactory agreement with this empirical reference data.
The behavior for high asymptotic $Sc$ is addressed by fitting an empirical scaling law,
\begin{equation}
  B_\theta(Sc) = A\,Sc^p\;,
  \label{eq:BT}
\end{equation}
for both available DNS \cite{Schwertfirm_Manhart:2007,Hasegawa_Kasagi:2009} and present ODT results.
The prefactor $A=10.5\pm0.01$ is similar in DNS and ODT, but the exponent $p$ is different, that is, $p\simeq0.72$ for DNS and $p\simeq0.67$ for ODT.
This difference in the parameterization roots in ODT's overestimation of $\theta_\tau$ which is related to the similarity of the momentum and scalar transfer (see below) and, hence, the limiting relation for the scalar diffusivity \cite{Shaw_Hanratty:1977,Klein_Schmidt_STAB:2021}.
For low asymptotic $Sc$ shown in figure~\ref{fig:tsca-fitCoef}(c), ODT predicts universal behavior for all $Re_\tau$ investigated, with $B_\theta\approx-10$ for CSV forcing. 
The scatter in the data seen for the lowest $Sc=0.025$ is due to a degraded fit quality that results from the poorly realized logarithmic region in the scalar profile $\Theta^+(y^+)$.

\subsection{\label{sec:Sct} Turbulent Schmidt number in the vicinity of the wall}

In this section, we diagnostically address the dissimilarity of the scalar and momentum transport for low and moderate $Sc$ with the aid of the turbulent Schmidt number,
\begin{equation}
  Sc_\text{t} = \dfrac{\nu_\text{t}}{\Gamma_\text{t}}\;,
 \label{eq:Sct}
\end{equation}
where $\nu_\text{t}=-\overline{u'v'}\big/ \left(\text{d}U/\text{d}y\right)$ and $\Gamma_\text{t}=-\overline{\theta'v'}\big/ \left(\text{d}\Theta/\text{d}y\right)$ denote the turbulent eddy viscosity and diffusivity, respectively. $\overline{u'v'}$ and $\overline{\theta'v'}$ are the wall-normal turbulent fluxes of the streamwise momentum and the scalar, respectively.
These fluxes are discussed in more detail below; their direct diagnostic computation within ODT is given in \ref{sec:cross-corr}.  

\begin{figure}[tp]
  \centering
  \includegraphics[height=52mm]{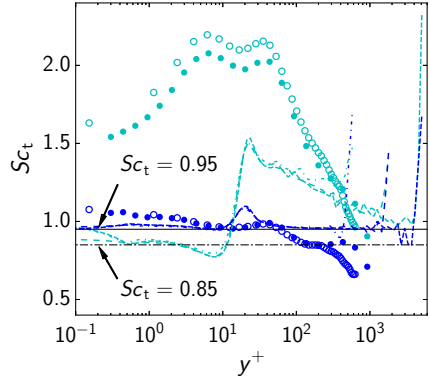} 
  \caption{%
    Near-wall structure of the turbulent Schmidt number $Sc_\text{t}$ for $Sc=0.71$ and $0.025$, respectively.
    ODT simulations with $Re_\tau=590$, $2000$, and $5200$ are for CSV, whereas reference DNS with $Re_\tau=640$ \cite{Abe_Kawamura_Matsuo:2004} (open symbols) and $1020$ \cite{Abe_Antonia:2019} (filled symbols) are for CSF. 
    Other line styles, symbols, and colors correspond with figures~\ref{fig:tmean}(a,\,c).
  }
  \label{fig:Sct}
\end{figure}

Figure~\ref{fig:Sct} shows wall-normal profiles of $Sc_\text{t}$ across the turbulent boundary layer.
ODT is generally closer to $Sc_\text{t}\simeq1$ than the reference DNS.
This is, in particular, the case for $Sc=0.71$ for which the ODT prediction $Sc_\text{t}\simeq0.95$ generally agrees with the near-wall value suggested by Abe and Antonia \cite{Abe_Antonia:2019}.
For $Sc=0.025$, ODT yields $Sc_\text{t}\approx0.85$ in the vicinity of the wall.
This differs from DNS \cite{Abe_Kawamura_Matsuo:2004,Abe_Antonia:2019}, which yields $Sc_\text{t}>1$ for $y^+<100$.
In the buffer layer, at $y^+\approx20$, ODT results exhibit an abrupt increase of $Sc_\text{t}$ to its maximum value of approximately 1.5 (approximately 2 for DNS \cite{Abe_Kawamura_Matsuo:2004,Abe_Antonia:2019}) from which $Sc_\text{t}$ gradually decreases to approximately 1 in ODT and DNS.
Even though ODT is unable to capture the near-wall structure of $Sc_\text{t}$ exactly, the model prediction is within physical bounds. 
Moreover, the model results for $Sc_\text{t}$ suggest that the momentum and scalar transport are indeed more similar in ODT than in DNS for finite $Re_\tau$ or $Pe_\tau$, respectively.

\subsection{\label{sec:trms} Scalar fluctuations and turbulent fluxes}

In addition to the mean field, ODT aims to resolve turbulent fluctuations and provide detailed flow statistics.
We therefore use ODT to estimate the variability of all flow variables without additional closure modeling.
As a first step in the fluctuation analysis, the variability and turbulent transport in the turbulent boundary layer is addressed below for the passive scalar.

\begin{figure}[tp]
  \centering
  \includegraphics[height=52mm]{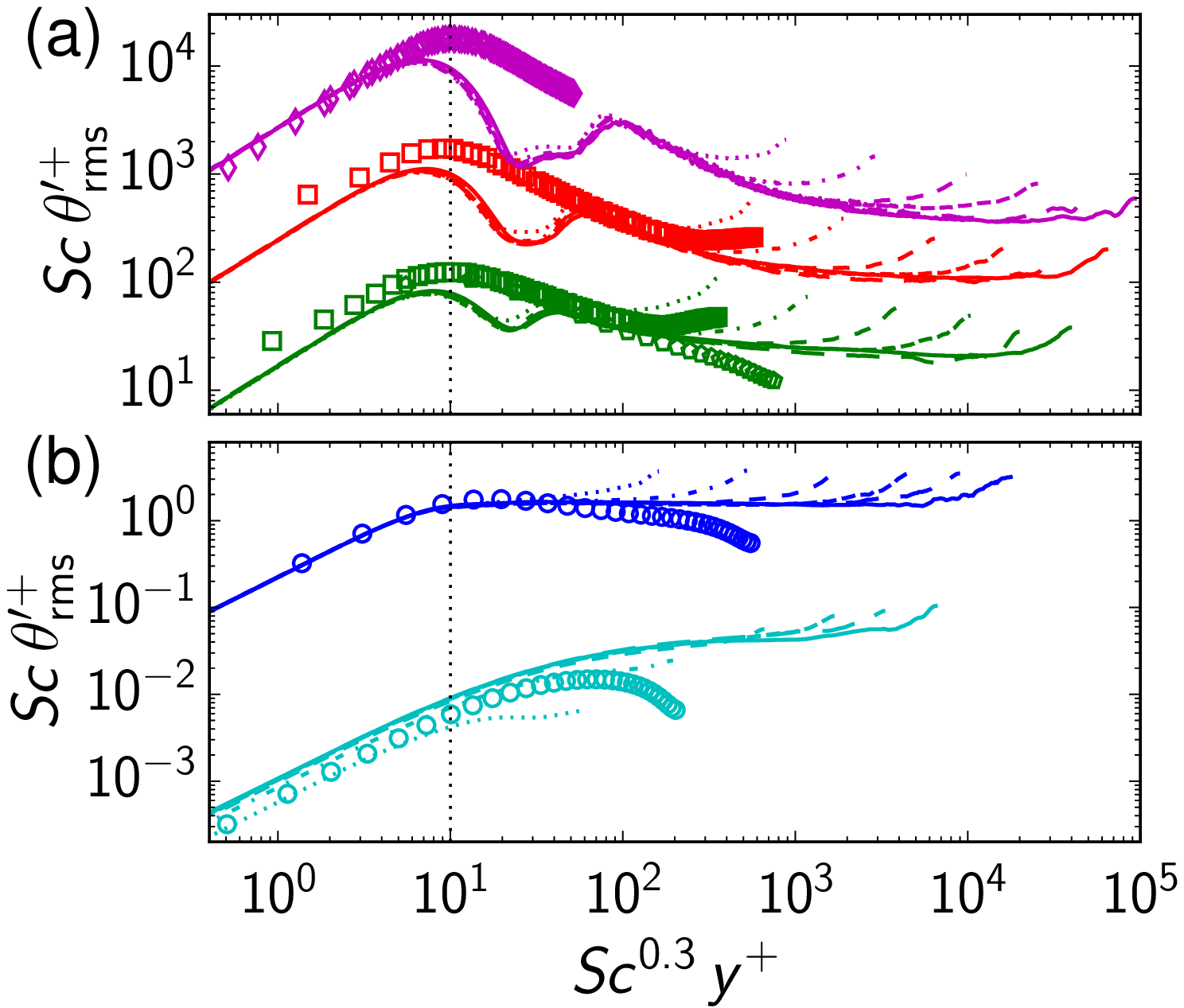} 
  \includegraphics[height=52mm]{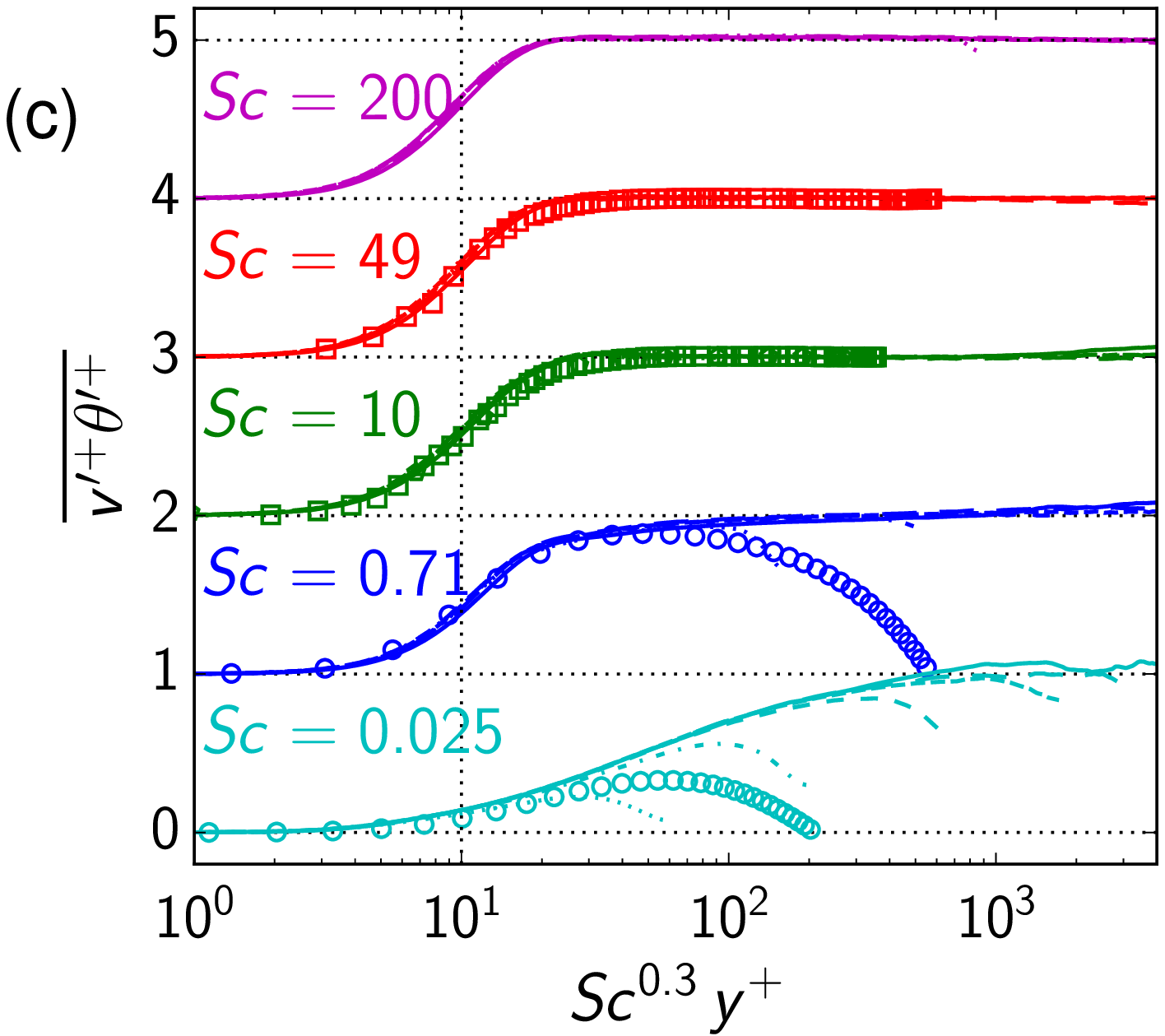}
  \caption{%
    (a,\,b) Root-mean-square (rms) scalar fluctuation, $\theta^{\prime +}_\text{rms}$, for high and low $Sc$, respectively, and
    (c) wall-normal turbulent scalar flux, $\overline{v^{\prime +}\theta^{\prime +}}$,
    for various $Sc$ and $Re_\tau$.
    Note that $\theta^{\prime +}_\text{rms}$ and $y^+$ are premultiplied with $Sc$ and $Sc^{0.3}$ for visibility and exploitation of scalar boundary layer similarity \cite{Schwertfirm_Manhart:2007}, respectively.    
    A dashed vertical line marks the nominal location of the high $Sc$ maximum root-mean-square fluctuation that corresponds to the inflection point in the turbulent flux.
    $\overline{v^{\prime +}\theta^{\prime +}}$ is vertically shifted by $1$ for visibility as indicated by dashed horizontal base lines.
    ODT results are for CSV only, whereas reference DNS are for both CSV (turning upward) and CSF (turning downward).
    Line styles, symbols, and colors as in figure~\ref{fig:tmean}.
  }
  \label{fig:trms}
\end{figure}

Figures~\ref{fig:trms}(a,\,b) show wall-normal profiles of the normalized root-mean-square scalar fluctuations, $\theta_\text{rms}^{\prime +}=(\overline{\theta^2} - \Theta^2)^{1/2} /\theta_\tau$, for various $Re_\tau$ and $Sc$ along with available reference DNS results.
These fluctuations notably increase with $Sc$ at small finite wall distance and towards the bulk for large $y^+$ in the case of CSV, whereas they reduce towards the bulk in the case of CSF forcing.
For $Sc=10$, reference data for CSV \cite{Schwertfirm_Manhart:2007} and CSF \cite{Abe_Kawamura_Matsuo:2004} exhibit perfect inner layer similarity.
This similarity is almost everywhere reproduced by ODT so that we limit our attention to CSV forcing for all other ODT cases shown.
Exceptional ODT behavior is limited to a finite but small wall-distance interval, $5<Sc^{0.3}\,y^+<100$, for large $Sc$.
For increasing $Sc>1$, a near-wall fluctuation peak develops at nominal location $Sc^{0.3}y^+\approx10$ \cite{Hasegawa_Kasagi:2009,Schwertfirm_Manhart:2007} that is indicated by a dotted line.
The peak width and location obey inner scaling arguments and are well-captured by ODT, but its shape and maximum are not.
The latter is a modeling artifact that is related to the triplet map \eqref{eq:triplet} used as an eddy micro-structure model to evaluate equations~\eqref{eq:eddy2} and \eqref{eq:eddyTau}.
This is discussed in \cite{Lignell_etal:2013} for the streamwise velocity fluctuations in terms of the inner $u_\text{rms}^{\prime +}$ peak that is shown below in figure~\ref{fig:vel}(c).
By model analogy, the artifact manifests itself also in the scalar fluctuations that are governed by the velocity boundary layer for $Sc>1$.
Hence, ODT aims to resolve transient wall-normal transport processes but is unable to resolve effects related to stream- and span-wise turbulent flow structures (e.g.~\cite{Hasegawa_Kasagi:2009}). 

Figure~\ref{fig:trms}(c) shows the turbulent scalar flux, $\overline{v^{\prime +}\theta^{\prime +}}$, which is estimated directly from the ensemble of discrete eddy events as described in \ref{sec:cross-corr}. 
ODT accurately captures the turbulent scalar flux for high asymptotic $Sc$ in agreement with reference DNS \cite{Schwertfirm_Manhart:2007} up to $Sc=49$.
The inflection point is located at $Sc^{0.3}y^+\approx10$, which is consistent with the location of the peak in $\theta_\text{rms}^{\prime +}$ even though ODT does not fully capture the latter peak up to $Sc=200$ shown.
The turbulent scalar flux approximately reaches $1$ at some wall distance irrespective of the forcing used for $Sc\geqslant0.71$ investigated.
This demonstrates surface layer similarity and dominance of turbulent transport at high $Sc$.
Towards the bulk, the turbulent scalar flux therefore depends on the scalar forcing used, that is, $\overline{v^{\prime +}\theta^{\prime +}}$ reaches a constant for CSV forcing, and turns down for CSF forcing.

For the low but finite $Sc=0.025$ investigated, the behavior is qualitatively different because scalar diffusion now heavily interacts with turbulent advection that enhances mixing across the turbulent boundary layer; this is addressed below in section~\ref{sec:tsr}.
Irrespective of the forcing used, the turbulent scalar flux reaches a local maximum and turns downward towards the bulk.  
ODT predictions and reference DNS \cite{Kawamura_Abe_Matsuo:1999,Abe_Kawamura_Matsuo:2004} are in agreement for $Re_\tau\geqslant180$ investigated, yielding a universal surface layer region $Sc^{0.3}y^+<20$ for which the turbulent fluxes of all reference and simulated cases collapse.

\subsection{\label{sec:jpdf} Probability density function of surface fluxes}

Conventional statistics discussed above have revealed that ODT is able to reasonably capture low-order statistical moments of the velocity and scalar boundary layer. 
This fundamental property is relevant, but might as well be achieved with conventional sub-filter scale closure modeling for the given application.
However, for modeling multiphysics boundary-layer flows it is also crucial to accurately capture detailed state-space statistical properties like correlations.
The ODT model has already been applied to such problems (e.g.~\cite{Monson_etal:2016,Medina_etal:2019,Kerstein_Wunsch:2006}).
Hence, wall-bounded scalar turbulence in channel flow serves as a canonical example in which the correlation of surface scalar and momentum fluxes has been documented in the literature.

Surface fluxes are resolved by the ODT model and they are obtained as time series of the fluctuating wall gradients of transported property fields. 
These wall gradients are notionally randomly perturbed by the turbulent boundary layer adjacent to the wall. 
Bulk quantities (like the bulk temperature in the case of CSF, or the turbulent drag that arises from the pressure drop across a channel segment) are very sensitive to the corresponding surface fluxes and their variability.
The relevant flux quantities are the instantaneous stream-wise wall-shear stress fluctuations, $\tau_{\text{w}}'=\tau_{\text{w}}-\bar{\tau}_{\text{w}}$, and the instantaneous surface scalar flux fluctuations, $q_\text{w}'=q_\text{w}-\bar{q}_\text{w}$.
These quantities are obtained from ODT simulations as an ensemble of time-series of wall gradients,
\refstepcounter{equation}
$$
  \tau_{\text{w}}' = \nu \frac{\partial u'}{\partial y}\bigg|_\text{w} \;,
  \quad
  q_\text{w}' = \Gamma \frac{\partial \theta'}{\partial y}\bigg|_\text{w} \;.
  \eqno{(\theequation{\mathit{a},\mathit{b}})}
  \label{eq:q-fluc}
$$

\begin{figure}[tp]
  \begin{subfigure}{0.45\textwidth}
    \caption{ODT $Sc=0.71$}
    \includegraphics[width=\linewidth]{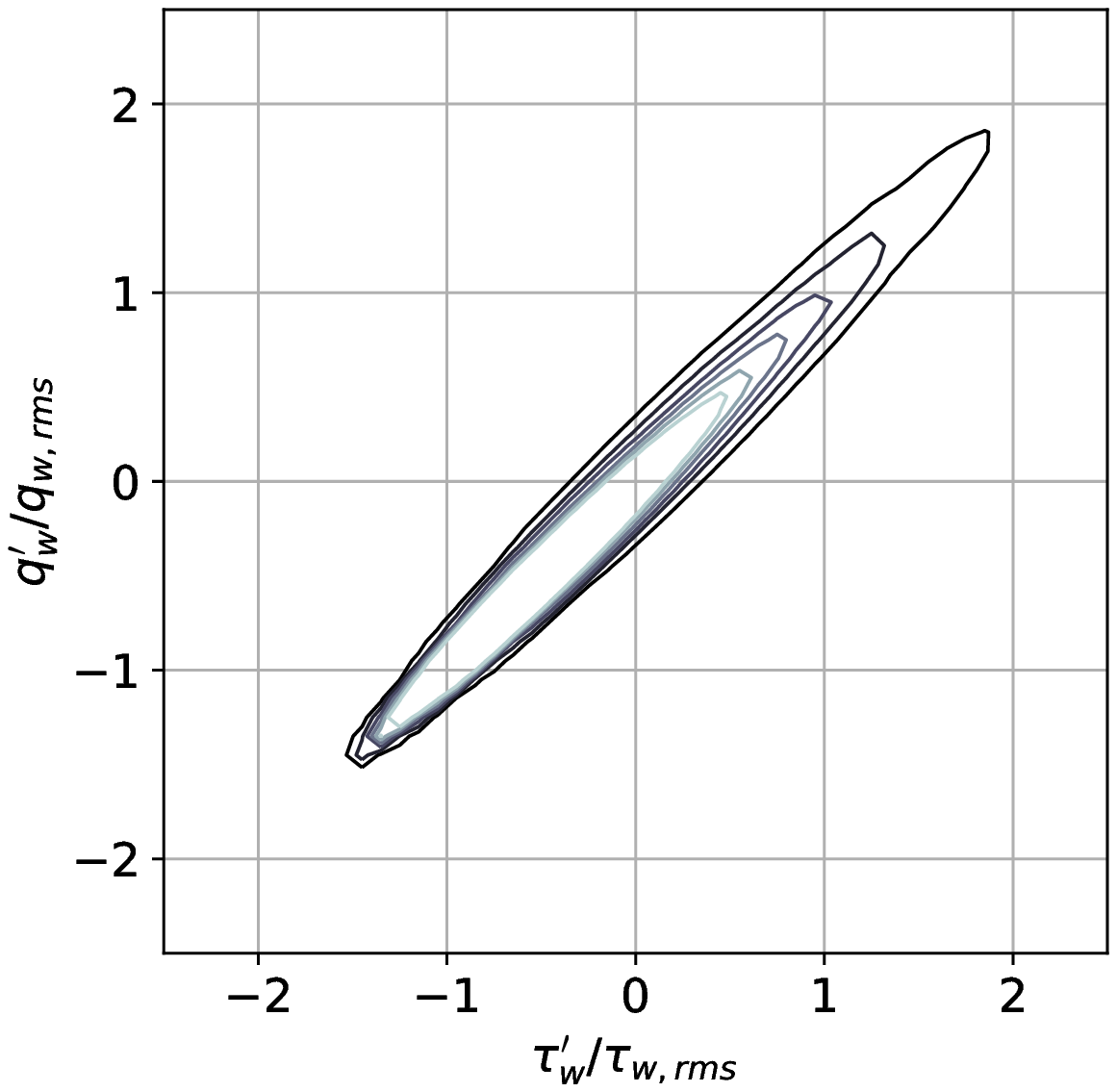}
  \end{subfigure}
  \begin{subfigure}{0.45\textwidth}
    \caption{DNS $Sc=0.71$}
    \includegraphics[width=\linewidth]{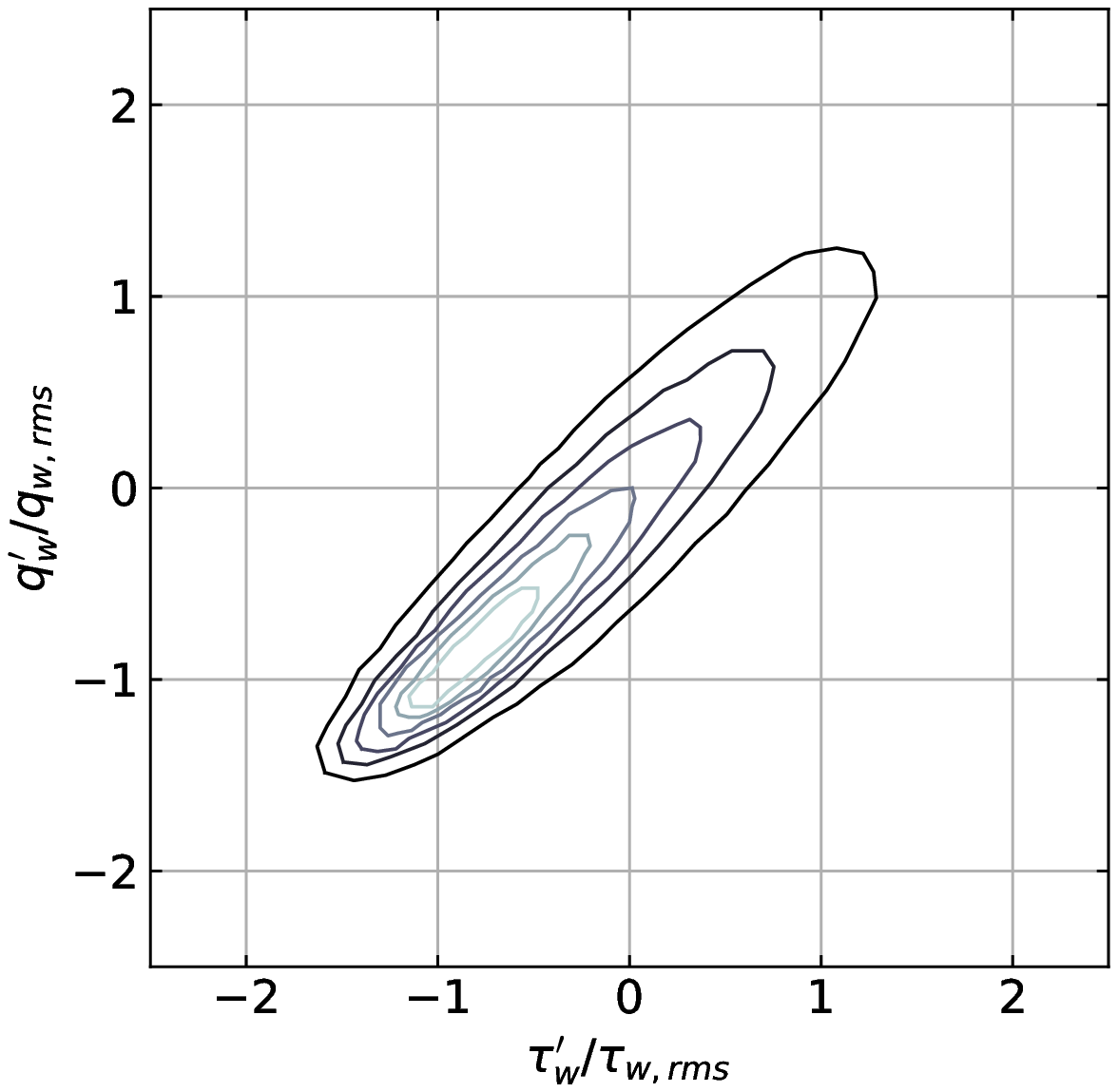}
  \end{subfigure}
  \\[2mm]
  \begin{subfigure}{0.45\textwidth}
    \caption{ODT $Sc=0.025$}
    \includegraphics[width=\linewidth]{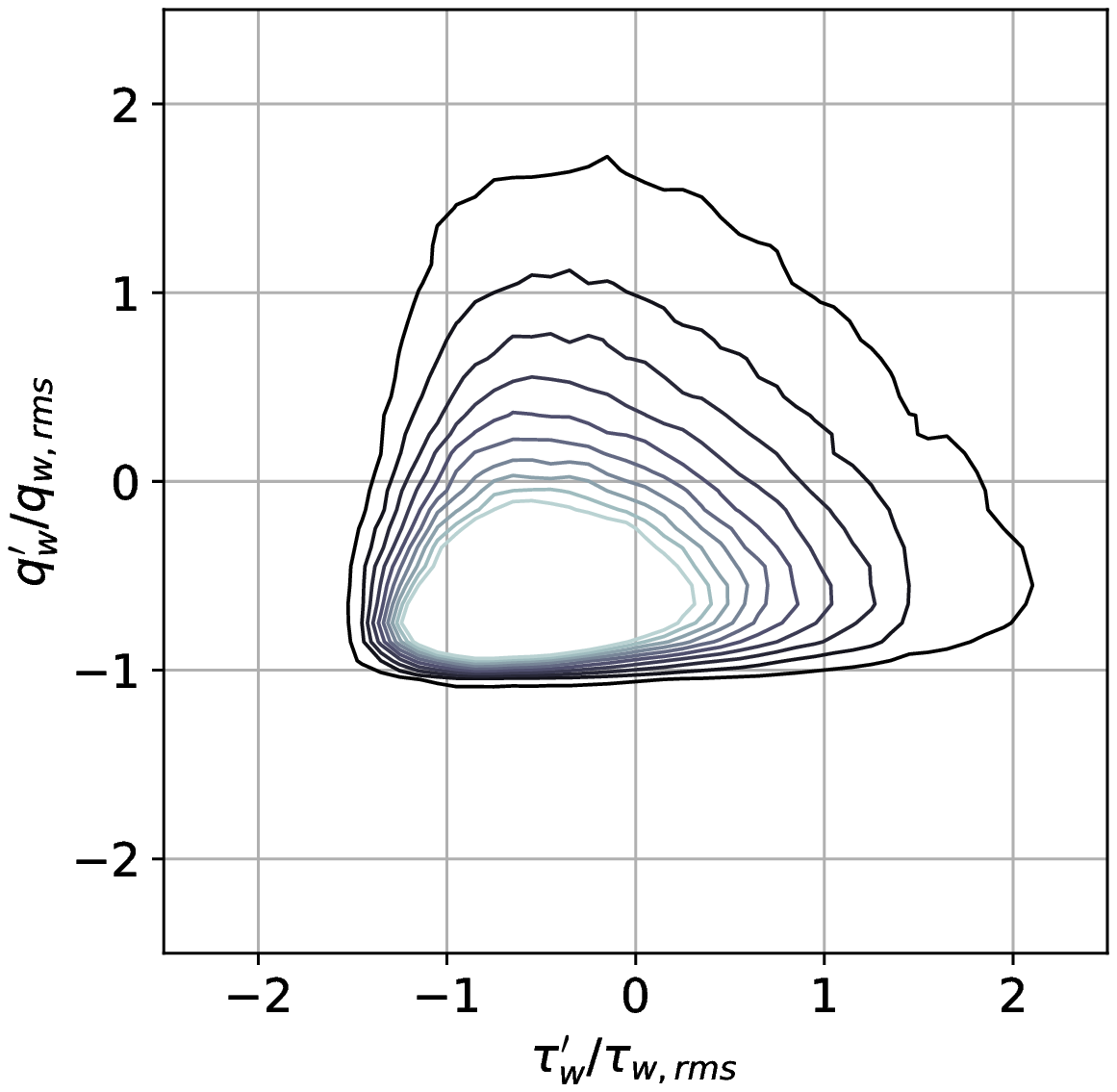}
  \end{subfigure}
  \begin{subfigure}{0.45\textwidth}
    \caption{DNS $Sc=0.025$}
    \includegraphics[width=\linewidth]{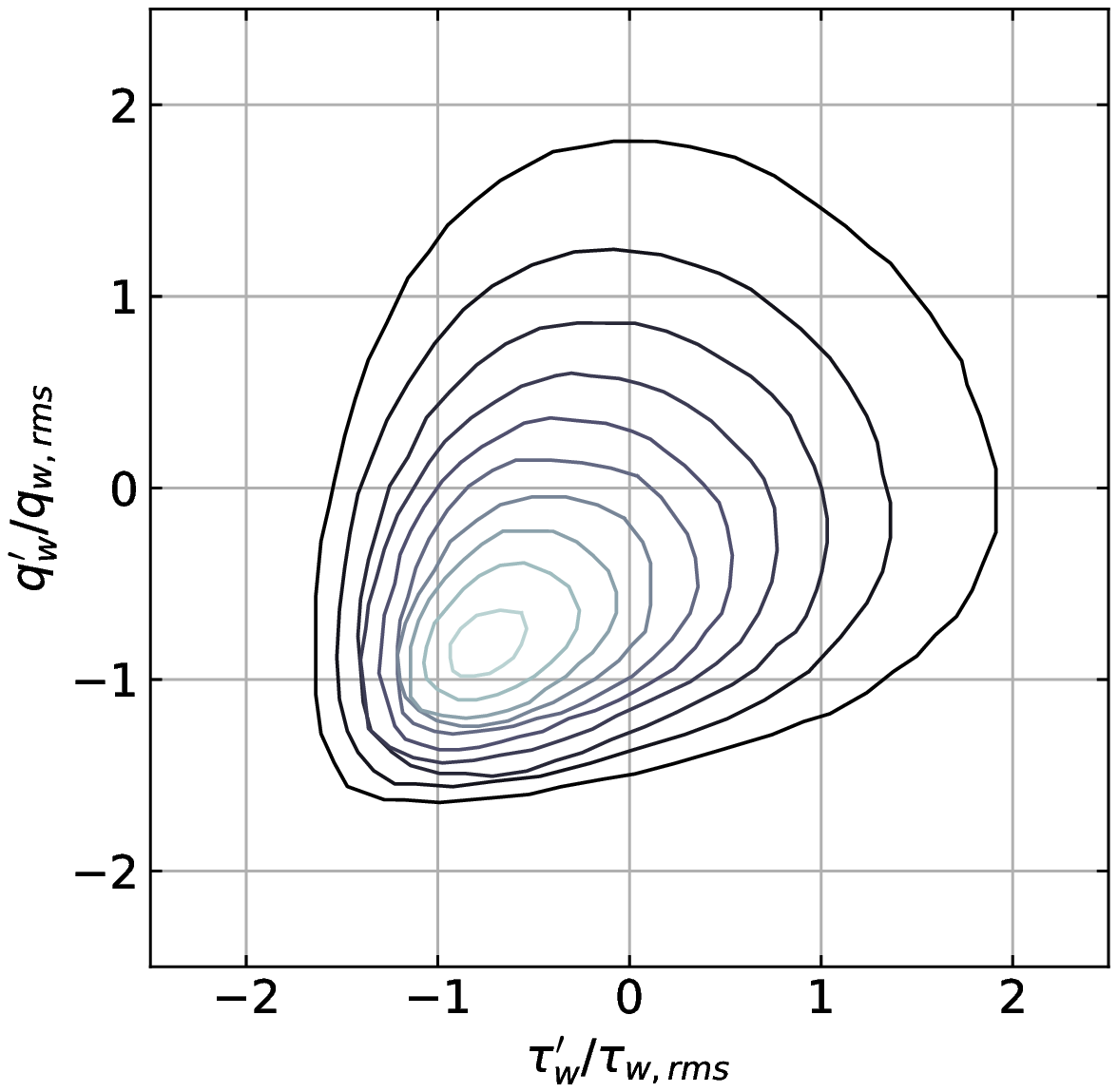}
  \end{subfigure}
  \caption{%
    Joint probability density functions (jPDFs) of the wall shear stress fluctuations, $\tau_\text{w}'$, and surface scalar flux fluctuations, $q_\text{w}'$, at the channel wall ($\text{w}$) for $Re_\tau=1000$ (CSF). 
    Fluctuating quantities are normalized by the respective root-mean-square (rms) values $\tau'_\text{w,rms}=\tau_\text{w,rms}$ and $q'_\text{w,rms}=q_\text{w,rms}$. 
    Probability density contours are shown starting at and spaced by $0.1$ for $Sc=0.71$, and starting at and spaced by $0.025$ for $Sc=0.025$. 
    Corresponding reference DNS results are from \cite{Abe_Kawamura_Matsuo:2004}.
  }
  \label{fig:jpdf} 
\end{figure}

Figure~\ref{fig:jpdf} shows joint probability density functions (jPDFs) of the instantaneous stream-wise wall-shear stress and the surface scalar-flux fluctuations for $Sc=0.71$ and $Sc=0.025$, respectively, for fully-developed turbulent channel flow with $Re_\tau=1000$ and CSF forcing of the scalar. 
Sixty bins were used in each direction over the indicated bounds in computing the jPDFs for ODT simulations.
ODT model predictions are shown in comparison with available reference DNS results \cite{Abe_Kawamura_Matsuo:2004}. 
The ODT model is able to capture relevant features of the reference DNS distribution functions for the same physical parameters and fixed model calibration.
The locations of the maxima and the spread around them are particularly well captured, which is consistent with the fidelity of the low-order statistical moments discussed above.

There are two notable discrepancies between the present ODT and reference DNS results.
One is the slightly narrower jPDF for moderate $Sc=0.71$ and more localized jPDF for small $Sc=0.025$, which indicates somewhat stronger correlation of the momentum and surface scalar fluxes in the model than in the DNS.
We attribute this to the near-wall triplet mapping in wall-attached eddy events that result in almost identical manipulations of the scalar and shear-dominating stream-wise velocity profiles there.

Another discrepancy is the missing negative surface scalar-flux contributions (below $-1$) for low $Sc=0.025$, which surprisingly is not an issue at $Sc=0.71$.
For small $Sc$, or, more precisely, small friction Peclet numbers, $Pe_\tau=Sc\,Re_\tau \lesssim1$, the scalar property field relaxes faster by molecular diffusion than it is stirred by turbulent eddies.
When the next wall-attached eddy occurs, the near-surface profile is already relaxed to a smooth monotonic profile.
Triplet mapping of such monotonic profiles can only increase the wall gradient but it can never reverse it.
The scalar wall gradient and, hence, the surface scalar-flux fluctuation, can not fall substantially below zero for $Pe_\tau\ll1$.
We emphasize that the low $Sc=0.025$ case is not constrained below $q_\text{w}'/q_\text{w,rms}=0$, it is constrained below $-1$. 
Hence, there is quite a bit of negative wall fluctuation.
The above argument implies that stronger negative scalar flux fluctuation in the DNS is due to the more complex three dimensional eddy structure \cite{Hasegawa_Kasagi:2009} that is not fully resolved by ODT.

Note that the wall fluxes are not actually negative, they are negative fluctuation about the mean.
Adopting inner scaling, equation \eqref{eq:nondim}, the ODT simulated mean and root-mean-square (rms) values are $\bar{\tau}_\text{w}^+=1$ and $\tau_\text{w,rms}^{\prime +}=0.363$, along with $\bar{q}_\text{w}^+=19.33$ ($19.60$) and $q_\text{w,rms}^{\prime +}=7.49$ ($7.81$) for $Sc=0.71$ ($0.025$).
The jPDF of the total surface fluxes can be obtained by shifting the origin by $\bar{\tau}_\text{w}/\tau_\text{w,rms}'=2.74$ and $\bar{q}_\text{w}/q_\text{w,rms}'=2.58$ ($2.51$), respectively.
Note further in this respect that the stream-wise wall-shear stress, that is, the total instantaneous momentum flux, may drop below zero due to the kernel mechanism, equations~(\ref{eq:eddy2},~\ref{eq:ci}), that acts in addition to the triplet mapping and models the effects of pressure-velocity correlations for any nonzero value of the model parameter $\alpha$.
Indeed, the extent of the negative viscous stress is very similar between the ODT and the DNS. 
However, negative stresses are extremely rare.
We emphasize that $Sc\neq1$ dictates the spread in the jPDF. 
In the singular case of $Sc=1$ (identical diffusivities) and $\alpha=0$ (only $u$ velocity is resolved) scalar and momentum transport are identical such that the jPDF will be an oblique line that is strictly bounded by nonzero total shear.

By contrast to what was just described, for moderate and large $Sc$, or more precisely, $Pe_\tau=Sc\,Re_\tau\geqslant1$, a cascade of eddy events can fit within the diffusion time scale. 
This means that a non-monotonic scalar profile due to a large eddy event is propagated to the wall by subsequent smaller events in the cascade.
This constitutes a mechanism for counter-gradient fluxes across the boundary layer that can lead to temporary negative wall gradients and surface fluxes. 
In addition, the described mechanism signalizes the relevance of rare large events for turbulent mixing processes.

While noting the differences in the jPDFs between the ODT and the DNS due to the mixing phenomenology of the ODT, and of course the lack of three-dimensional flow structures in the ODT, the overall qualitative agreement is quite remarkable given the relative simplicity of the ODT simulations.  
These results serve to highlight the importance and sensitivity of physical processes occurring in the flow that can be difficult to isolate with a single model that is constrained in its representation of the flow (even if such constraint is physically realistic, as in the case of DNS).

\subsection{\label{sec:tvar} Scalar variance budget}

The scalar variance budget is primarily useful to elucidate the missing scalar fluctuations at high $Sc$ (see figure~\ref{fig:trms}(a) above). 
But, in general, the scalar variance budget will help us to obtain a quantitative understanding of the fluctuating passive scalar transport in comparison to the TKE (see figure~\ref{fig:tke}). 
The scalar variance transport equation that defines the budget of terms is given by (see e.g.~\cite{Schwertfirm_Manhart:2007})
\refstepcounter{equation}
$$
  \frac{\partial \overline{\theta^{\prime2}}}{\partial t}
   + \boldsymbol{U} \boldsymbol{\cdot} \boldsymbol{\nabla} \overline{\theta^{\prime2}}
   = P_\theta + \varepsilon_\theta + D_\theta + T_\theta \;,
  \eqno{(\theequation)}
  \label{eq:tvar}
$$
where $\overline{\theta^{\prime2}}=\theta_\text{rms}^{\prime2}$ denotes the scalar variance and $P_\theta$ the production, $\varepsilon_\theta$ the dissipation, $D_\theta$ the diffusive transport, and $T_\theta$ the turbulent transport of the scalar variance.
The left-hand side vanishes for statistically stationary channel flow, and the terms on the right-hand side only retain their wall-normal contributions within the ODT modeling framework, that is,
\begin{subequations}
 \renewcommand{\theequation}{\theparentequation \textit{\alph{equation}}}
 \label{eq:tvar-odt}
 \begin{align}
  P_\theta &= -2\overline{v'\theta'} \,\frac{\text{d} \Theta}{\text{d} y} \;,
  \\
  \varepsilon_\theta &= -2\Gamma \overline{ \frac{\partial \theta'}{\partial y}   \frac{\partial \theta'}{\partial y} } \;,
  \\
  D_\theta &= \Gamma \frac{\text{d}^2 \overline{\theta^{\prime2}}}{\text{d} y^2} \;,
  \\
  T_\theta &= -\frac{\text{d} \,\overline{v'\theta^{\prime2}}}{\text{d} y} \;.
 \end{align}
\end{subequations}
The corresponding dimensionless expressions for these terms are obtained by division with $\theta_\tau^2 u_\tau^2/\nu$, which is indicated by the super-script `$+$'.

\begin{figure}[tp]
  \centering
  \includegraphics[height=52mm]{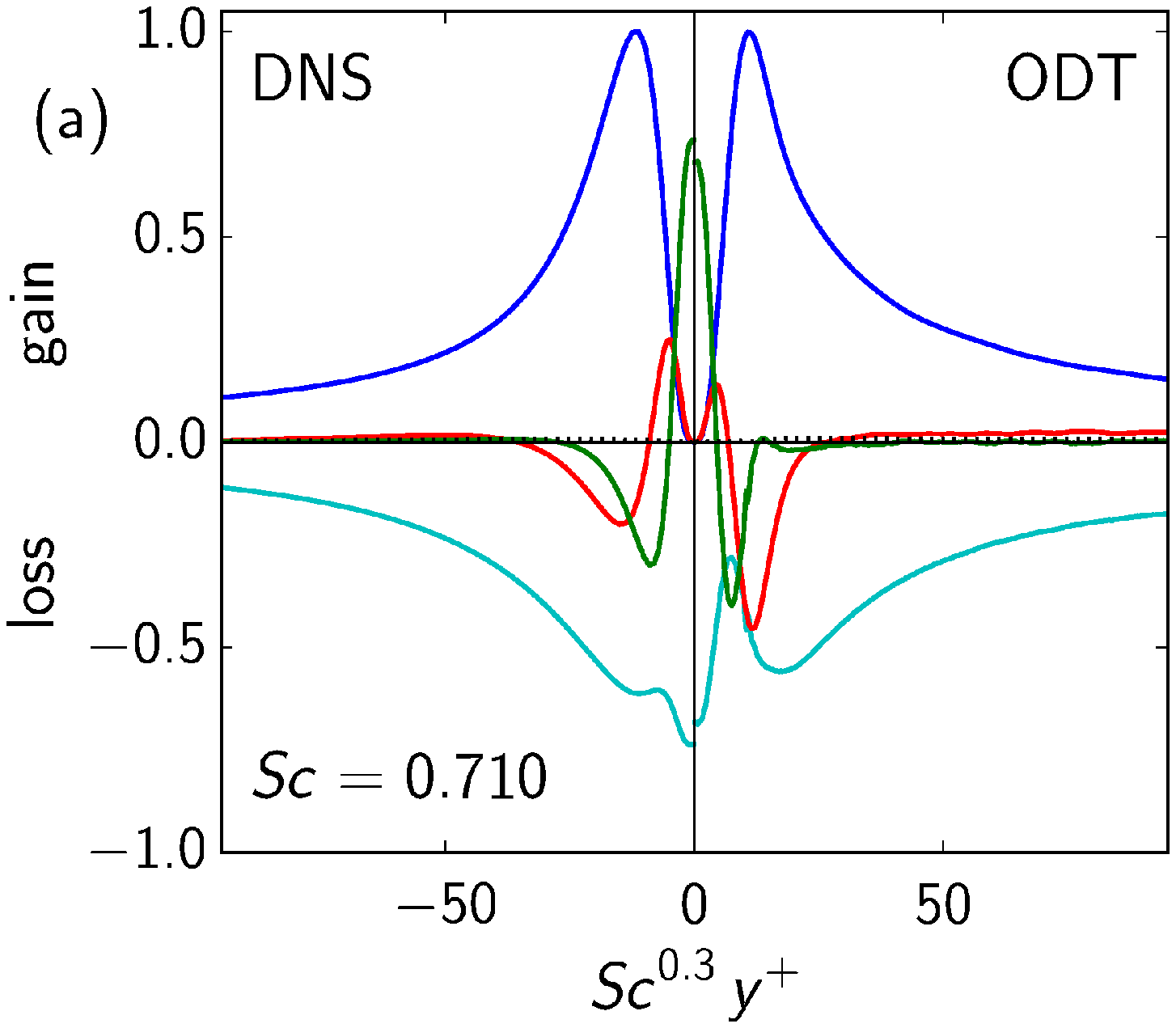} 
  \includegraphics[height=52mm]{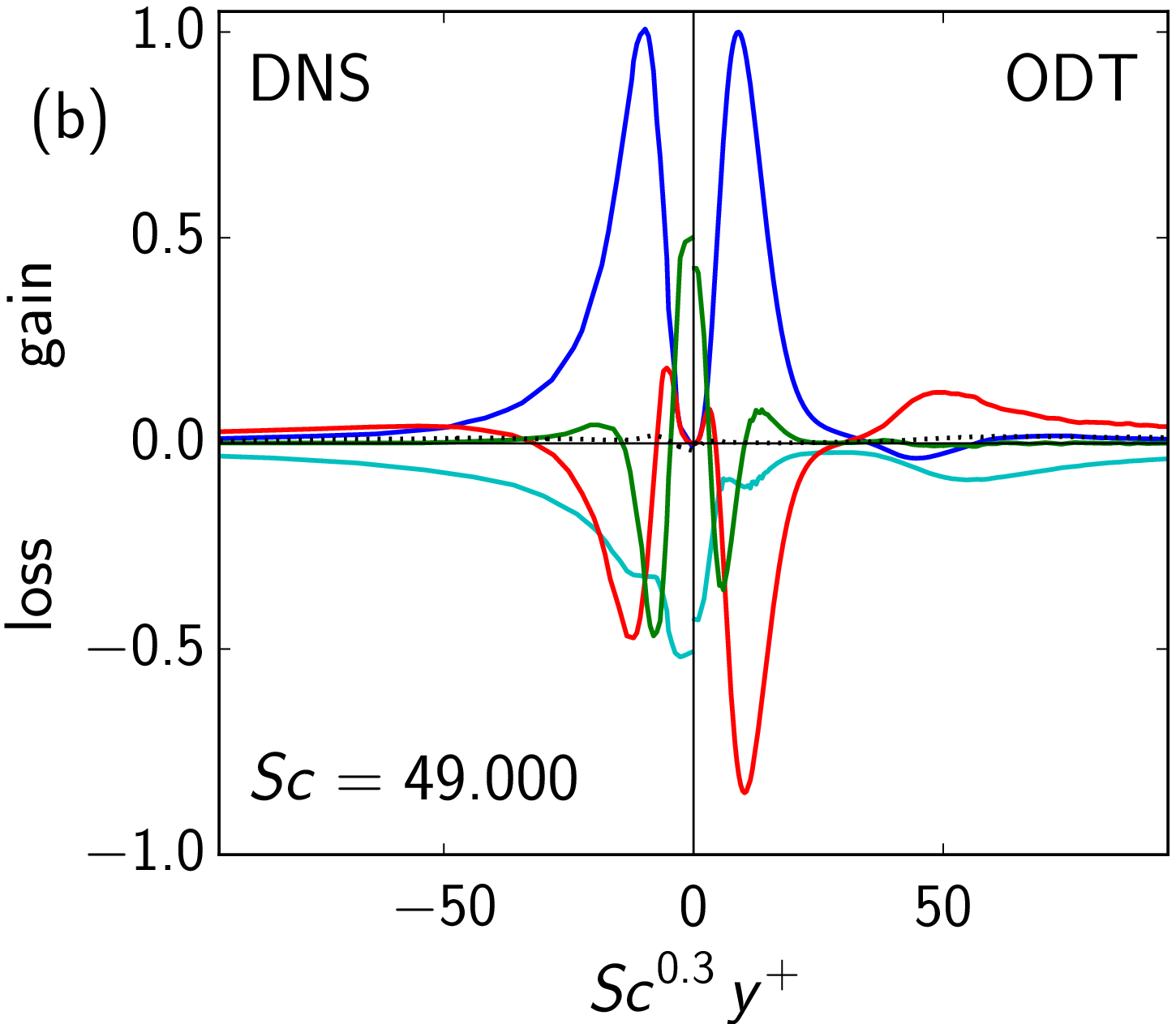}
  \includegraphics[height=52mm]{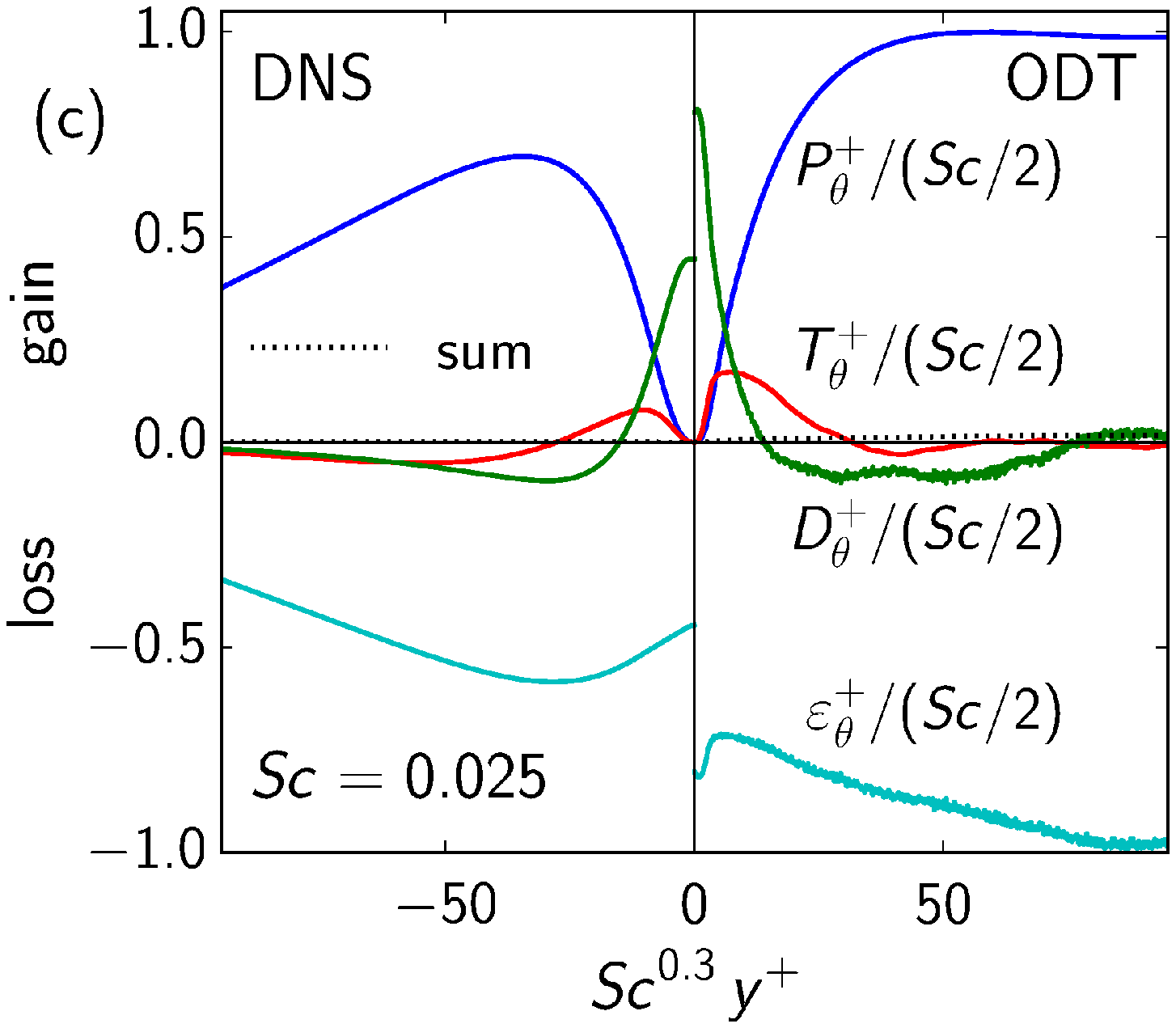} 
  \caption{%
    `Back-to-back' plot of the scalar variance budget balance given by equation~\eqref{eq:tvar-odt} for ODT ($y^+\geqslant0$) and equation~\eqref{eq:tvar} for reference DNS ($y^+\leqslant0$).
    (a)~$Sc=0.71$, ODT: $Re_\tau=590$ (CSV), DNS: $Re_\tau=640$ (CSF)~\cite{Kawamura_Abe_Matsuo:1999};
    (b)~$Sc=49$, ODT: $Re_\tau=180$ (CSV), DNS: $Re_\tau=180$ (CSV)~\cite{Schwertfirm_Manhart:2007}; and
    (c)~$Sc=0.025$, ODT: $Re_\tau=590$ (CSV), DNS: $Re_\tau=640$ (CSF)~\cite{Kawamura_Abe_Matsuo:1999}.
    Colors distinguish the contributions by $P_\theta$, $T_\theta$, $D_\theta$, and $\epsilon_\theta$, respectively.
  }
  \label{fig:tvar}
\end{figure}

Figures~\ref{fig:tvar}(a--c) show the normalized scalar variance budget balance for both ODT and reference DNS for $Sc=0.71$, $Sc=49$, and $Sc=0.025$, respectively.
The dimensionless terms are additionally normalized for each case with the corresponding peak production, $P^+_{\theta,\text{max}}=Sc/2$ (e.g. \cite{Schwertfirm_Manhart:2007}).
ODT exhibits a balance of the scalar variance budget as the right-hand-side terms of equation~\ref{eq:tvar} sum to zero.
The gross structure of all terms is captured by ODT, but somewhat better for the moderate $Sc=0.71$ than for the higher $Sc=49$, and generally less well for low $Sc=0.025$.

For moderate and high $Sc$, figures~\ref{fig:tvar}(a,b), the turbulent production peak is correctly predicted with some discrepancy in its tail as the $Sc$ number increases.
At the same time, dissipation is increasingly underestimated so that compensating transport maintains the balance.
The compensation is primarily achieved by turbulent transport and less by diffusive transport.
This is consistent with the previously discussed modeling artifacts in the root-mean-square fluctuation for large $Sc$.

For low $Sc$, figure~\ref{fig:tvar}(c), ODT still exhibits a balance, but now both production and dissipation are overestimated in magnitude whereas the turbulent and diffusive variance fluxes are reasonably well captured.
We attribute the former to the triplet mapping \eqref{eq:triplet} in the discrete eddy events.
Small-scale scalar variance is artificially enhanced by instantaneous mappings that introduce profile gradient discontinuities that are quickly regularized by deterministic molecular diffusion due to which the scalar variance dissipation is also artificially increased. 
This reveals the relevance of resolving the time-scale separation between turbulent advection and molecular diffusion.
Interestingly, this is a local effect that is almost balanced due to time-scale separation between turbulent advective and molecular diffusive transport processes for $Sc\ll1$.
This separation is resolved by ODT such that the turbulent fluxes of the scalar fluctuation variance are well captured by the model, albeit production and dissipation are not.
Below, we address ODT's capabilities for capturing fluid-surface interactions of turbulent mixing in wall-bounded flows. 

\subsection{\label{sec:tsr} Mixing time scale}

Physical processes in boundary-layer flows are sensitive not only to near-wall fluctuations and turbulent transport but also to the local mixing efficiency.
In particular, this is the case for flows that exhibit intense scalar-wall interaction (like wall fires \cite{Monson_etal:2016} or scalar mixing processes in the atmospheric surface layer \cite{Owinoh_etal:2005}). 
In the following, we therefore address ODT's capabilities for capturing molecular mixing processes and fluctuation damping for various independent $Sc$ scalars in the vicinity of a wall under turbulent flow conditions.
The related mixing efficiency is quantified by the mixing-time-scale ratio, $R$, that is given by
\begin{equation}
  R = \frac{\overline{\theta^{\prime2}}/\varepsilon_\theta}{2k/\varepsilon} \;,
  \label{eq:tsr}
\end{equation}
where $k$ denotes the turbulence kinetic energy (TKE) and $\varepsilon$ the TKE dissipation in the lower-order model as defined in equation~\eqref{eq:tke-odt}.

\begin{figure}[tp]
  \centering
  \includegraphics[height=52mm]{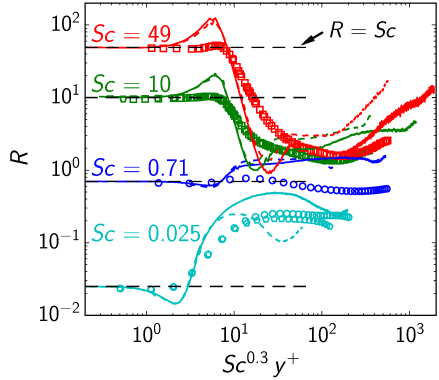}
  \caption{%
    Mixing-time-scale ratio, $R$, from equation~\eqref{eq:tsr} across the turbulent boundary layer for various $Sc\leqslant49$ and $Re_\tau\leqslant640$, which has been limited only to aid visibility.
    Reference DNS results are from \cite{Abe_Kawamura_Matsuo:2004,Kawamura_Abe_Matsuo:1999,Schwertfirm_Manhart:2007}.
    Line styles, symbols, and colors as in figure~\ref{fig:tmean}.
  }
  \label{fig:tsr}
\end{figure}

Figure~\ref{fig:tsr} shows wall-normal profiles of the mixing-time-scale ratio, $R$, by comparing ODT predictions with reference DNS and both are found to exhibit reasonable agreement.
Due to inner layer similarity discussed above in sections~\ref{sec:tmean}, \ref{sec:trms}, and \ref{sec:tvar}, only ODT results with CSV forcing are shown for low $Re_\tau$ that correspond with available reference DNS.
The vicinity of the wall is dominated by molecular diffusion which is why $R\to Sc$ for $y^+\to0$ for any finite $Sc$ and $Re_\tau$.
With increasing distance from the wall, that is, for $Sc^{0.3}y^+>10$, we find $R\simeq 1$ which is consistent with the analysis of the turbulent boundary-layer structure (section~\ref{sec:tmean}) that the scalar and momentum transport tend to be more similar towards the bulk.
Across the log layer, both scalar and momentum (velocity) mixing is dominated by the inertial range of the turbulence cascade.
For even larger distances from the wall, specifics of the outer layer and bulk flow influence the mixing behavior.
Note in this regard that the diffusion-dominated scalar surface layer thickness decreases like $Sc^{-0.3}$ (e.g.~\cite{Shaw_Hanratty:1977,Hasegawa_Kasagi:2009}) such that it becomes negligibly small for $Sc\to\infty$ in relation to the viscous sublayer with thickness $y^+\approx5$ (e.g.~\cite{Pope:2000}).
The scalar transfer and mixing properties are then dominated by the momentum boundary layer (e.g.~\cite{Shaw_Hanratty:1977,Kader_Yaglom:1972}) which is captured by ODT.

\subsection{\label{sec:Sh} Sherwood number and scalar transfer coefficient}

We have shown above that ODT resolves transient transport processes in the whole boundary layer including surface flux fluctuations.
An important remaining question that we address in this section is: how well can ODT capture and predict the scalar transfer to a wall?

Scaling regimes of the scalar transfer to a wall are quantified by the Sherwood number, $Sh\sim q/q_\Gamma$, which expresses the total scalar flux, $q$, in units of the purely molecular diffusive flux, $q_\Gamma$, under (assumed) absence of turbulence and up to a multiplicative constant.
The Sherwood number is the analog of the Nusselt number in the context of mass rather than heat transfer that is here generalized as scalar transfer.
When temperature variations are small and internal heating due to viscous dissipation negligible (which is often justifiable), both Sherwood and Nusselt numbers are isomorphic, in particular with respect to their scaling laws.
In this study, either $q$ is directly prescribed by $q_\text{w}$ in the case of isoflux wall-boundary conditions (CSF forcing) or it is obtained as model result in the case of Dirichlet wall-boundary conditions (CSV forcing).
For CSV forcing, that is considered in more detail in the following, $q_\Gamma$ is proportional to the prescribed bulk-wall (half wall-to-wall) scalar difference, $\mathrm{\Delta}\theta = |\theta_\text{top}-\theta_\text{bot}|/2$, yielding $q_\Gamma=\Gamma\,\mathrm{\Delta}\theta/\delta$ for channel flow as sketched in figure~\ref{fig:config}(a).

For fully-developed channel flow, the total scalar flux is carried by molecular diffusion such that $Sh$ is related to the scalar transfer coefficient, $K^+$, as (e.g.~\cite{Kader_Yaglom:1972,Wei_etal:2005}) \refstepcounter{equation}
$$
  Sh = \gamma\,Re_\tau\,Sc\, K^+(Re_\tau,Sc)\;, \qquad
  K^+ = \dfrac{\theta_\tau}{\mathrm{\Delta}\theta} \;,
  \eqno{(\theequation{\mathit{a},\mathit{b}})}
  \label{eq:Nu}
$$
in which $\gamma$ denotes a conventional geometrical proportionality constant and $\theta_\tau$ the friction scalar property defined in equation~(\ref{eq:nondim}\textit{b}).
Following \cite{Schwertfirm_Manhart:2007}, we select $\gamma=d_\text{h}/(2\delta)=2$, which implies $Sh$ based on the hydraulic diameter, $d_\text{h}=4\delta$, rather than the channel half-height $\delta$.
A conventional Nusselt number, therefore, is $Nu=Sh$ (e.g.~\cite{Kawamura_etal:1998,Abe_Kawamura_Matsuo:2004,Pirozzoli_etal:2016}) that exhibits the same $Sc$ and $Re_\tau$ dependence as $Sh$. 

\begin{figure}[tp]
  \centering
  \includegraphics[height=52mm]{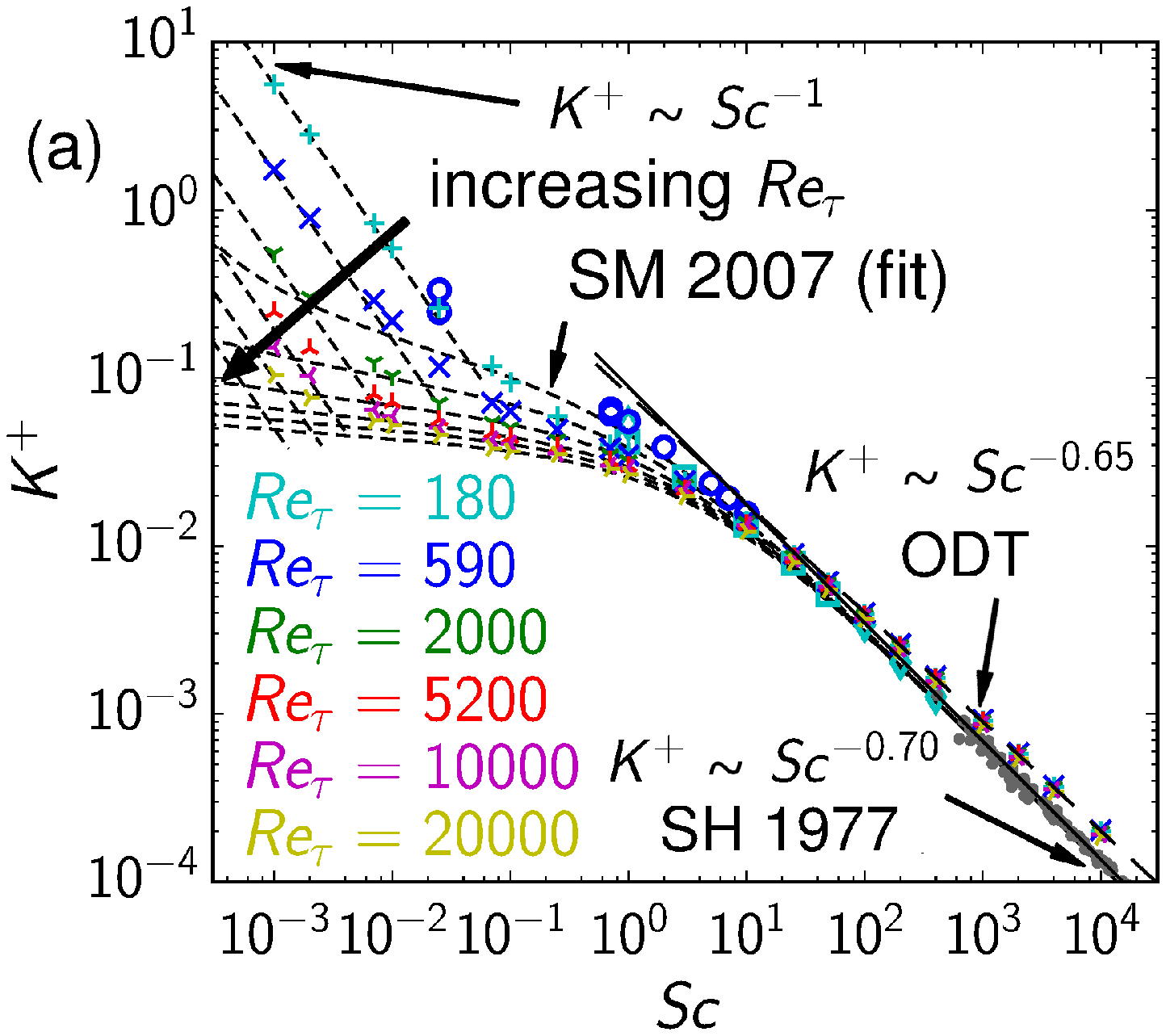} 
  \includegraphics[height=52mm]{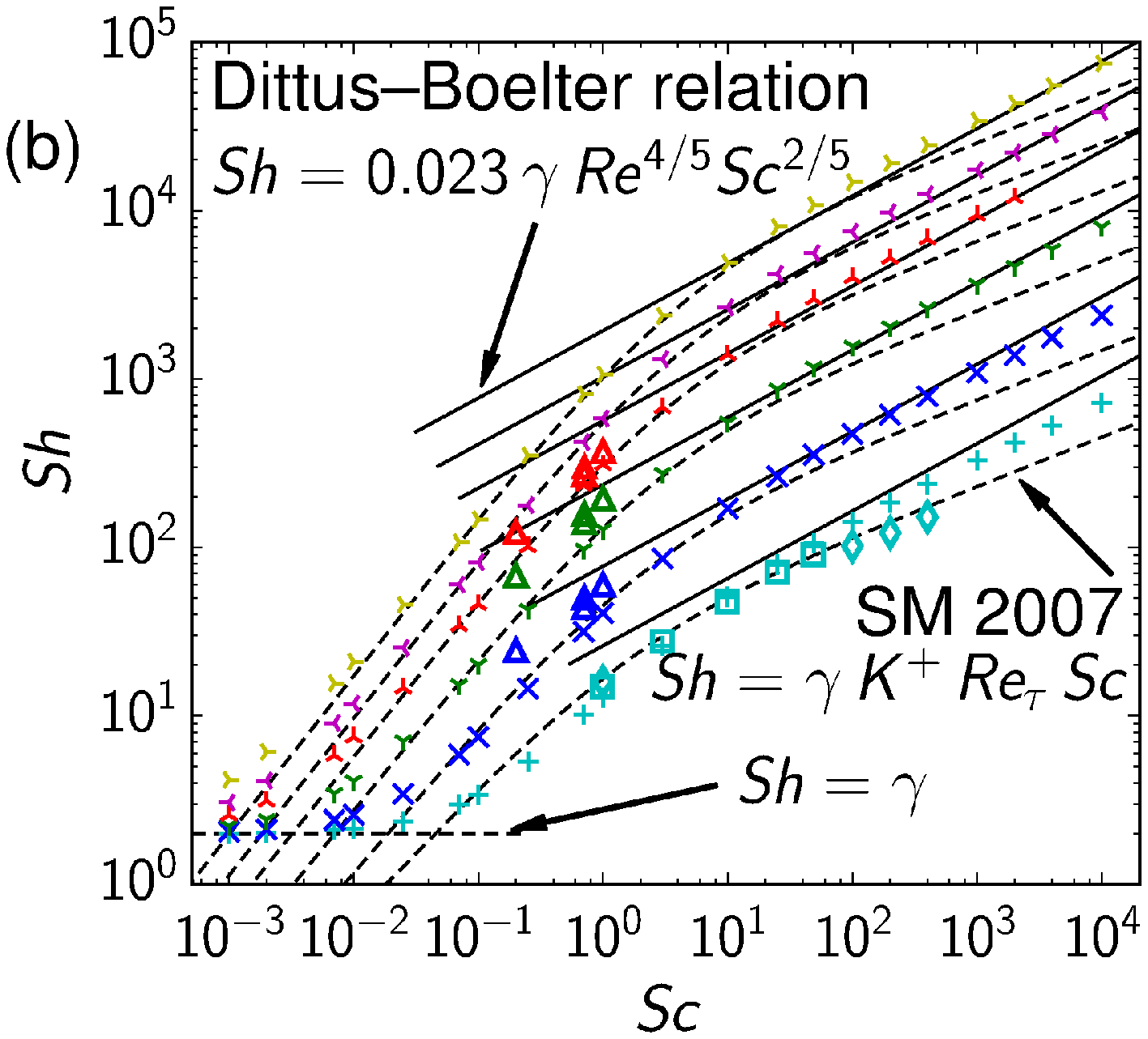} 
  \\
  \includegraphics[height=52mm]{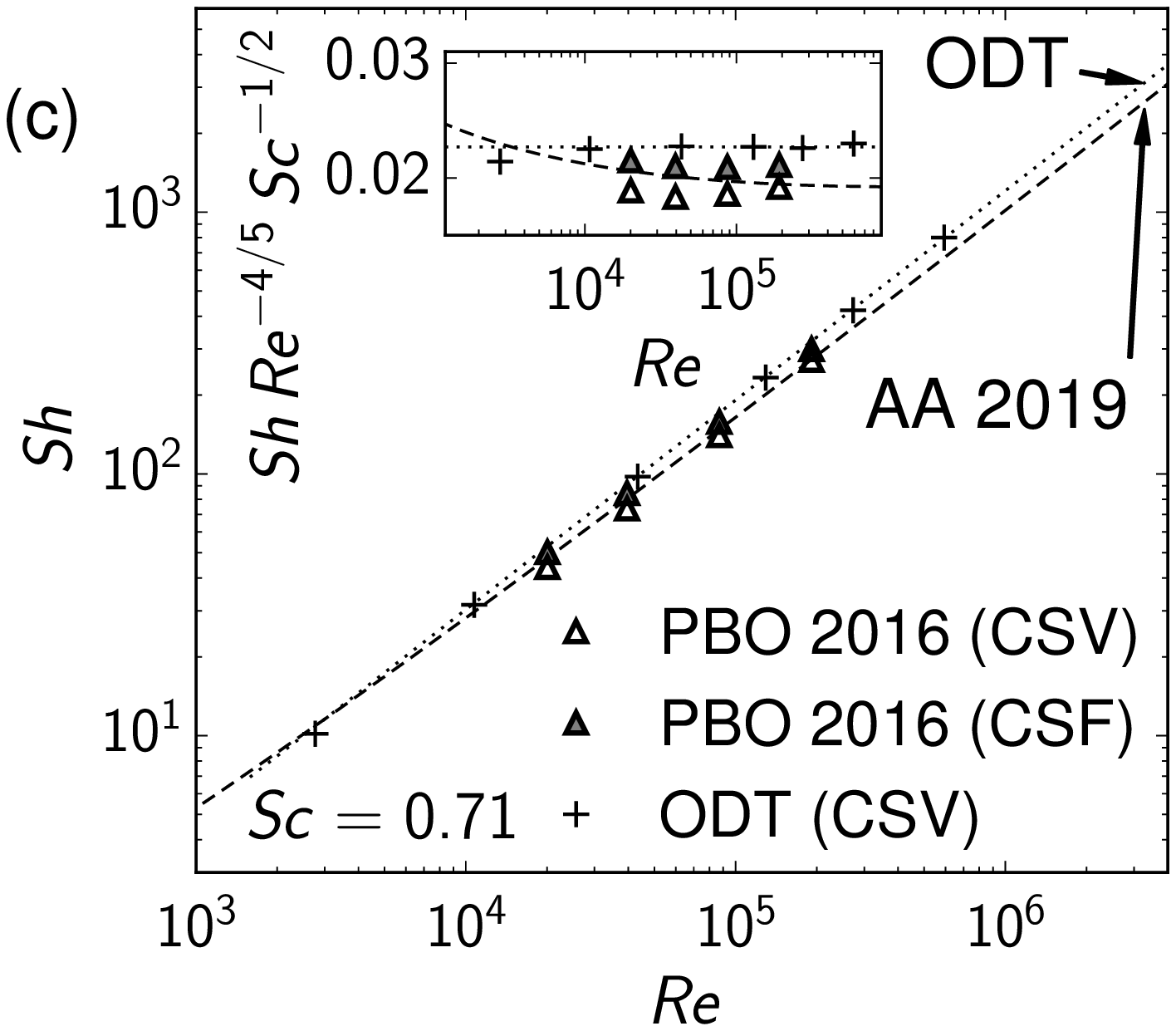}
  \includegraphics[height=52mm]{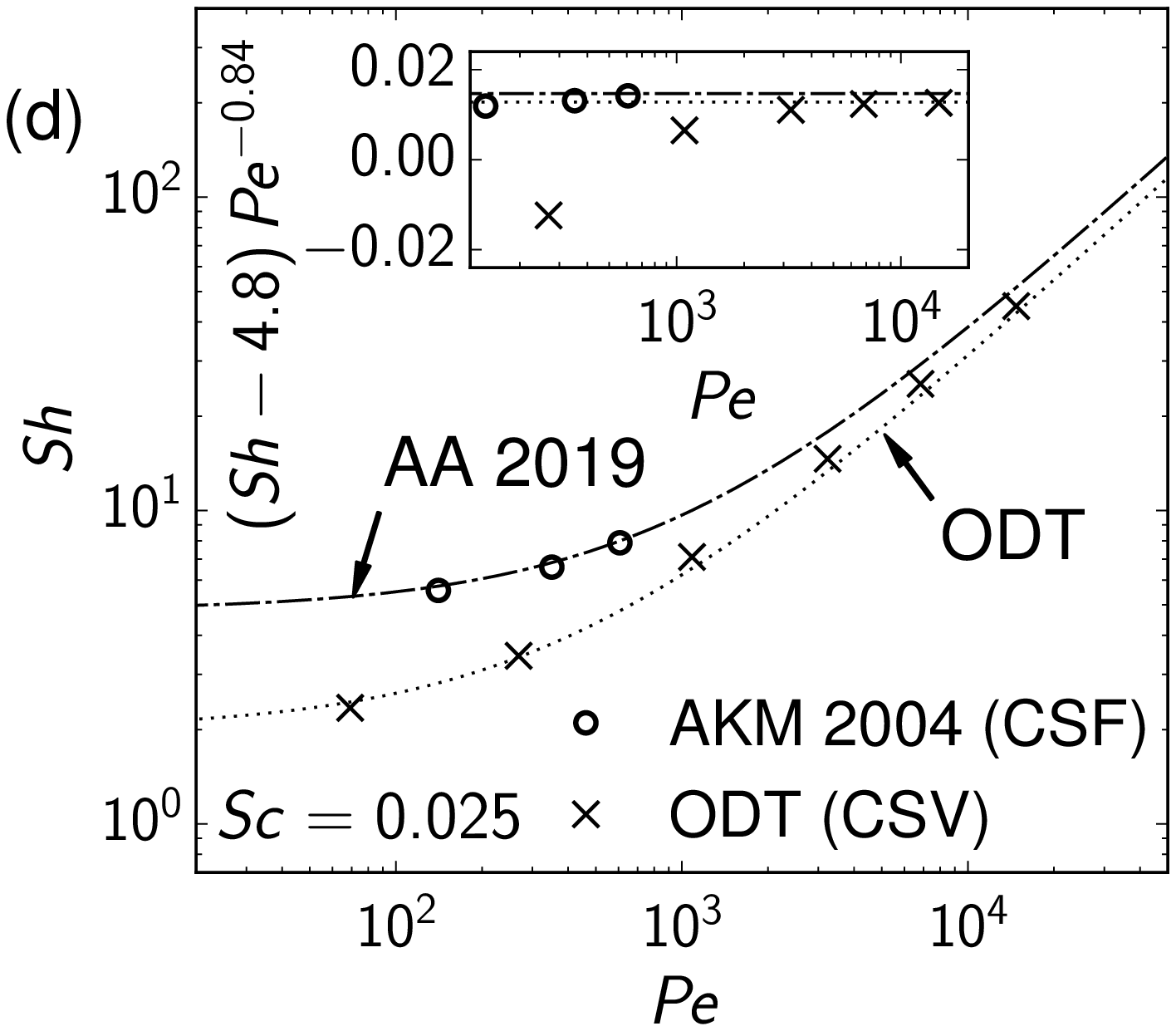}
  \caption{%
    (a)~Scalar transfer coefficient, $K^+$, and (b)~Sherwood number, $Sh$, as functions of the Schmidt number, $Sc$, for various $Re_\tau$.
    Reference empirical scaling relations for $K^+$ and $Sh$ are based on equations~\eqref{eq:Nu}, \eqref{eq:Nu_DB} \cite{Dittus_Boelter:1985}, and \eqref{eq:Nu_SM} \cite{Schwertfirm_Manhart:2007}.
    Symbols and references as in figure~\ref{fig:tmean}, but here colors in (a,\,b) distinguish $Re_\tau$.
    Additional high $Sc$ reference experiments (grey dots) are from \cite{Shaw_Hanratty:1977}. 
    (c)~$Sh$ as function of the bulk Reynolds number, $Re$, for $Sc=0.71$ comparing ODT, DNS \cite{Pirozzoli_etal:2016}, and the scaling due to the expressions collected in equation~\eqref{eq:NuRe} \cite{Abe_Antonia:2019}.
    The inset shows the same data but compensated based on equation~\eqref{eq:Nu_KC}.
    (d)~$Sh$ as function of the bulk Peclet number, $Pe$, for $Sc=0.025$ comparing ODT, DNS \cite{Abe_Kawamura_Matsuo:2004}, as well as equation~\eqref{eq:NuPe} for DNS \cite{Abe_Antonia:2019} and equation~\eqref{eq:NuPe_ODT} for ODT.
    The inset shows the same data but compensated based on equation~\eqref{eq:NuPe}.
  }
  \label{fig:Sh}
\end{figure}

Figures~\ref{fig:Sh}(a,\,b) show $K^+$ and $Sh$, respectively, for various $Sc$ passive scalars in turbulent channel flows at different $Re_\tau$.
ODT predictions are shown together with available reference DNS, pipe flow measurements (filled gray bullets) \cite{Shaw_Hanratty:1977}, and empirical scaling relations.
Three different scaling regimes of the scalar transfer can be discerned, corresponding to low, intermediate, and high $Sc$.
ODT results exhibit very good to reasonable agreement with the reference data and scaling relations that seems to improve for high asymptotic $Re_\tau$.
We proceed in the following by addressing the scaling regimes individually.  

We first discuss the diffusive limit, $Sh\to\gamma=2$ for $Sc\to0$, that exhibits an increasing time-scale separation that makes numerical simulations costly or barely feasible for applications such as heat transfer in liquid metal batteries (e.g.~\cite{Kolesnichenko_etal:2020}).
ODT can accommodate large time-scale separation.  
For constant $Sh$, equations~\eqref{eq:nondim} and \eqref{eq:Nu} yield the presence of a purely diffusive scalar profile.
For CSV forcing, this profile is steady and linear yielding $K^+\to Sc^{-1}\,Re_\tau^{-1}$. 
Hence, $Sh$ becomes parametrically independent of $Re_\tau$ for asymptotically small $Sc$.
This is fulfilled in practice for any small friction Peclet number, $Pe_\tau=Sc\,Re_\tau\ll1$.
The latter condition limits the maximum permissible friction Reynolds number to $Re_\tau\ll Sc^{-1}$ for diffusion-dominated scalar transfer. 

We now turn to the high $Sc$ and $Re_\tau$ flow regime that is relevant for various applications but numerically inaccessible due to high computational resource requirements that are imposed by the Kolmogorov and Batchelor scales as well as the sampling time (e.g.~\cite{Schwertfirm_Manhart:2007,Ostilla-Monico_etal:2015,Pirozzoli_etal:2016}).
A lower-order model like ODT makes small-scale resolving numerical simulations feasible.
The model aims to resolve fluctuating wall-normal molecular and turbulent transport processes on all relevant scales such that the mean scalar transfer is a model result. 
Reference empirical scaling relations are available that were obtained by measurements and theoretical considerations (e.g.~\cite{Kader_Yaglom:1972,Sleicher_Rouse:1975,Wei_etal:2005,vanReeuwijk_Lari:2012}).
Some widely used scaling relations parameterize the $Sc$ and $Re$ dependence of $Sh$ by a power law,
\begin{equation}
  Sh\simeq0.023\,\gamma\,Re^{4/5}\,Sc^{b} \;,
  \label{eq:Nu_DB}
\end{equation}
where $b=2/5$ yields the Dittus--Boelter relation \cite{Dittus_Boelter:1985}, $b=1/3$ the well-established Colburn relation (e.g.~\cite{McAdams:1954,Shaw_Hanratty:1977}), and $b=1/2$ a relation that is closer to the Prandtl number dependence of convective heat transfer (e.g.~\cite{Kays_Crawford:1980}).
Note that $Re=U_\text{b}\delta/\nu$ denotes the bulk Reynolds number, where $U_\text{b}=(2\delta)^{-1}\int_0^{2\delta} U(y)\,\text{d}y$ is the bulk velocity, which is here conveniently obtained from the mean velocity profile $U(y)$. 
For fully-developed turbulent channel flow, $Re$ can be obtained from the empirical relation $Re_\tau\simeq0.18\,Re^{0.88}\sim Re^{7/8}$ \cite{Pope:2000}.
Both $Re$ and $Re_\tau$ are given in table~\ref{tab:config} for the present ODT simulations.
Those data confirm that ODT exhibits the same scaling exponent, but the prefactor is $\approx0.16$ due to underestimation of the velocity in the outer layer (see figure~\ref{fig:vel}(a)).
The ODT results shown in figure~\ref{fig:Sh}(b) for $Sc\gtrsim10$ fall between the Dittus--Boelter and Colburn relations with a tendency towards the former.
This is consistent with the ODT limiting relation $K^+\sim Sc^{-0.65}$ \cite{Klein_Schmidt_TSFP:2017,Klein_Schmidt_STAB:2021} that is slightly different from the reference $K^+\sim Sc^{-0.70}$ \cite{Shaw_Hanratty:1977,Schwertfirm_Manhart:2007,Hasegawa_Kasagi:2009}.

It is apparent by now that finite $Sc$ and $Re$ effects play a role for the intermediate scaling regime.
These effects were already addressed by Sleicher and Rouse \cite{Sleicher_Rouse:1975}, who provided an empirical scaling relation together with the applicability region as
\begin{align}
  Sh &= 5 + 0.15\,Re^a\,Sc^b 
  \label{eq:Nu_SR}
  \\
  &\text{with}\quad
  a = 0.88 - \dfrac{0.24}{4 + Sc}\;, \quad
  b = \dfrac{1}{3} + \dfrac{\text{e}^{-0.6\,Sc}}{2}
  \notag \\
  &\text{for}\quad
  Sc\in\left[10^{-1}, 10^4\right], \quad
  Re\in\left[10^4, 10^6\right].
  \notag
\end{align}
The above relation is given for completeness of the discussion since it has been widely used to address finite $Sc$ effects (e.g.~\cite{Kawamura_etal:1998}).
Here, we do \emph{not} show equation~\eqref{eq:Nu_SR} explicitly in the figures because of the following two reasons.
First, $Sh$ given in equation~\eqref{eq:Nu_SR} reaches the physically incorrect limit $Sh\to5$ for $Sc\to0$, which is already notable for moderately low $Sc\simeq10^{-1}$.
Second, the above parameterization yields $Sh\sim Sc^{1/3}$ for $Sc\to\infty$, which is simply the Colburn scaling that is readily included in a more generally applicable and physically based parameterization that is discussed next.

Schwertfirm and Manhart~\cite{Schwertfirm_Manhart:2007} used DNS in conjunction with turbulent boundary-layer theory to yield the semi-empirical scaling
\begin{align}
 K^+ 
  &= \dfrac{Sh}{\gamma\,Re_\tau\,Sc}
   = \dfrac{\kappa_\theta}{\ln Re_\tau + \kappa_\theta\,\xi\,Sc^{1-r} + r\,\ln Sc - \ln \xi} 
 \label{eq:Nu_SM}
 \\
 &\text{with}\quad
 \kappa_\theta=0.27\;, \quad \xi=11.5\;, \quad r=0.29
 \quad\text{for}\quad Sc\gtrsim1 \;.
 \notag
\end{align}
In this equation, the logarithmic correction emerges due to the presence of scalar inner linear and log layers.
The analysis put forward by \cite{Schwertfirm_Manhart:2007} is strictly valid only for $Sc\gtrsim1$ such that the linear and log layer scalar profiles intersect as can be seen in figure~\ref{fig:tmean}(a,\,b) for $Sc\geqslant0.71$.
This intersection unambiguously defines a diffusive scalar boundary-layer thickness, $\delta_\Gamma^+$, which, for asymptotically high $Sc$, takes the form $\delta_\Gamma^+\simeq\xi\,Sc^{-r}$, where $r\approx0.3$ \cite{Schwertfirm_Manhart:2007,Shaw_Hanratty:1977}.
Note that the value of $r\approx0.3$ corresponds with the similarity scaling of the scalar boundary layer coordinate, $Sc^{0.3}\,y^+$, that was used in various figures above.
Note further that $r$ in the asymptotic relation for $\delta_\Gamma^+$ implies that $Sh$ due to equation~\eqref{eq:Nu_SM} approaches the Colburn scaling, $Sh\sim Sc^r\approx Sc^{1/3}$ (e.g.~\cite{McAdams:1954}), for high asymptotic $Sc$.
Also note that we keep the parameterization coefficients $\kappa_\theta$, $\xi$, and $r$ unchanged here for clarity.
Reparameterization using ODT results is possible but yields similar results \cite{Klein_Schmidt_STAB:2021}.

Relation~\eqref{eq:Nu_SM} is shown in figure~\ref{fig:Sh}(a,\,b) with thin curved dashed lines that serve for orientation and as bound on the scalar transfer for extrapolated $Sc<1$.
To the best of our knowledge, no generalized and physically based relation is currently available for this regime.
(Equation~\eqref{eq:Nu_SR}, for example, fails to capture the limiting behavior for $Sc\lesssim10^{-1}$ as discussed above.)
The fitting parameters are from \cite{Schwertfirm_Manhart:2007} and were obtained with the aid of DNS for turbulent channel flow at $Re_\tau=180$ with $Sc<50$ passive scalars using CSV forcing.
This parameterization predicts $K^+$ for another DNS at $Re_\tau=150$ with $Sc\leqslant400$ \cite{Hasegawa_Kasagi:2009} and  measurements in pipes at high asymptotic $Re_\tau$ and $Sc$ of the order $10^3$ to $10^4$ \cite{Shaw_Hanratty:1977}.
ODT captures this behavior but would yield somewhat different numerical fitting parameters due to the marginally captured emerging dissimilarity of the near-surface scalar and momentum transport \cite{Klein_Schmidt_TSFP:2017,Klein_Schmidt_STAB:2021}.
The origin of this effect is discussed above in sections~\ref{sec:tmean}, \ref{sec:jpdf}, and \ref{sec:tvar}, showing that it is related to the resolved and unresolved model physics.
We note that, for finite $Re_\tau$ and $Sc$, ODT results are well described by relation~\eqref{eq:Nu_SM}.
This implies that local effective scalings, $Sh\sim Sc^b$, with $b>2/5$ are realized that are steeper than the Dittus--Boelter and Colburn relations \cite{Tschisgale_Kempe:2021}.
This demonstrates the model's capabilities for capturing regime transitions in scalar transfer by resolving a relevant fraction of the state space of surface-flux fluctuations. 
Next, we investigate the bulk Reynolds and Peclet number dependence, respectively, for fixed $Sc$ in order to quantify the influence of the momentum transfer.

Figure~\ref{fig:Sh}(c) shows $Sh$ as a function of the bulk Reynolds number $Re$ for $Sc=0.71$ that corresponds to the case of heat transfer in air.
The reference scaling shown in the figure is from Abe and Antonia \cite{Abe_Antonia:2019}.
This scaling is derived by assuming presence of fully-turbulent velocity and scalar boundary layers, application of the Reynolds analogy, $C_\text{f}=2K_\text{t}$, for $Sc=O(1)$, and utilization of the skin-friction log law such that
\begin{subequations}
  \renewcommand{\theequation}{\theparentequation \textit{\alph{equation}}}
  \label{eq:NuRe}
  \begin{align}
    Sh &= K_\text{t} \, Re\, Sc \;, 
    \\
    K_\text{t} &= \dfrac{ \sqrt{C_\text{f}/2} }{ 2.18\,\ln\big( (Re/2) \,\sqrt{C_\text{f}/2} \big) + 2.40 } \;, 
    \\
    C_\text{f} &= \dfrac{\tau_\text{w}}{\rho U_\text{b}^2/2} = 2 \dfrac{Re_\tau^2}{Re^2} \;,
  \end{align}
\end{subequations}
where $K_\text{t}$ denotes the bulk scalar transfer and $C_\text{f}$ the skin friction coefficient, respectively.
For finite $Re$, equation~(\ref{eq:NuRe}\textit{a}) yields effective power-law scalings $Sh\sim Re^p$, where $p(Re)$ becomes approximately constant, $p\approx4/5$, for high asymptotic $Re$, as used in equation~\eqref{eq:Nu_DB}.
A frequently used parameterization for internal flow is \cite{Kays_Crawford:1980}
\begin{equation}
  Sh\simeq0.021\,Re^{4/5}\,Sc^{1/2} \;,
  \label{eq:Nu_KC}
\end{equation}
which is used for compensating the high $Re$ data in the inset of figure~\ref{fig:Sh}(c).
Available reference DNS \cite{Pirozzoli_etal:2016} data agrees with the theoretical scaling given by equations~(\ref{eq:NuRe}).
This scaling predicts slightly lower $Sh$ than the empirical relation~(\ref{eq:Nu_KC}).
ODT exhibits satisfactory agreement with the reference scaling and DNS \cite{Pirozzoli_etal:2016} for all $Re>1000$ investigated.
The model correctly predicts $Sh\sim Re^{4/5}$ for high asymptotic Reynolds numbers, but it is unable to capture finite $Re$ effects and exhibits a slightly larger prefactor of $0.0227$ (dotted line).
This is not just a model calibration issue but is intimately related to the outer layer velocity deficit explained above.

Figure~\ref{fig:Sh}(d) shows $Sh$ as function of the bulk Peclet number, $Pe=Sc\,Re$, for $Sc=0.025$, that corresponds to the case of heat transfer in mercury.
Abe and Antonia \cite{Abe_Antonia:2019} developed a reference scaling that has been parameterized as
\begin{equation}
  Sh\simeq4.8 + 0.0147\,Pe^{0.84} 
  \quad \text{for}\quad Pe\leqslant1000 \;,
  \label{eq:NuPe}
\end{equation}  
where $Pe$ should not be too small since the relation does not extrapolate to the diffusive limit, which is $Sh\to Sh_0=\gamma=2$ for $Pe\to0$.
Available reference DNS \cite{Abe_Kawamura_Matsuo:2004} for low $Pe$ agree with the simple relation given in equation~(\ref{eq:NuPe}), but the $Pe$ (or $Re$) numbers reached for $Sc=0.025$ are too small to assess the high $Pe$ asymptotic behavior.
This is possible, however, with ODT that accurately predicts the empirical power-law scaling $Sh\sim Pe^{0.84}$ for $Pe>1000$ investigated, although it systematically underestimates $Sh$ by a constant offset since the modeling error reduces with increasing $Pe$ (see inset of figure~\ref{fig:Sh}(d)).
A simple parameterization (dotted line) that extrapolates from the diffusive limit, $Sh_0=2$, to high asymptotic $Pe$ and that approximately describes present ODT results is
\begin{equation}
  Sh\simeq2 + 0.0147\,Pe^{0.84} \;.
  \label{eq:NuPe_ODT}
\end{equation}
This relation has not previously been reported and it is very similar to the reference scaling given in equation~(\ref{eq:NuPe}) except for the additive constant.
Even though the model prediction for low $Pe\leqslant1000$ differs notably from the reference data \cite{Abe_Kawamura_Matsuo:2004,Abe_Antonia:2019}, equation~(\ref{eq:NuPe_ODT}) nevertheless demonstrates that ODT obeys fundamental physical bounds of the scalar transfer. 
The model, therefore, has good predictive capabilities in the high $Pe\geqslant O(1000)$ flow regime that is challenging for, or not accessible to, DNS for the foreseeable future (e.g.~\cite{Alcantara-Avila_etal:2018,Abe_Antonia:2019}).

\section{\label{sec:conc} Conclusion}

One-dimensional turbulence (ODT) numerical simulations of passive scalars turbulent channel flow have been performed up to $Re_\tau=20{,}000$ for $Sc\in\left[10^{-3},10^4\right]$ with the scalar prescribed either by constant scalar value (CSV) or constant scalar flux (CSF) wall-boundary conditions.
These simulations were made feasible by utilizing a fully-adaptive ODT implementation as stand-alone tool that aims to resolve transient wall-normal transport processes on all relevant flow scales within a lower-order stochastic framework.
The free model parameters were calibrated once for the velocity boundary layer such that the $Sc$ and $Re_\tau$ dependencies reported are model predictions.
ODT consistently predicts low-order flow statistics throughout the turbulent boundary layer with perfect surface (inner) layer similarity for high asymptotic Reynolds numbers irrespective of the scalar forcing used.
Wall-normal fluxes are very well captured, but the model fails to resolve some fluctuations that are related to stream- and span-wise flow structures.
This modeling error manifests itself by an unphysically degraded local near-wall scalar fluctuation maximum and an overestimation of the scalar wall-gradient for CSV and $Sc\gg1$.
Nevertheless, the ODT solutions obey inner scaling and collapse for all $Re_\tau$ investigated.
Joint probability density functions (jPDFs) of the fluctuating wall-shear stress and surface scalar flux reveal that the negative scalar flux fluctuations are more constrained in ODT than in DNS.
This is another manifestation of the dimensional modeling error since the three-dimensional eddy structure in the DNS that can support the negative scalar flux fluctuations, but not the map-based advection model that is used in ODT.
The consequence of this modeling error is a generally more similar (but not identical) scalar and momentum transport, in particular across the logarithmic layer.
Nevertheless, ODT is able to accurately capture local mixing time scales which is a crucial property for application to multiphysics boundary layers.
Finally, it was shown that the model is able to accurately predict the scalar transfer to the surface which is quantified by the Sherwood (Nusselt) number, $Sh(Sc,Re_\tau)$.
Present ODT results reproduce the scaling relation proposed by Schwertfirm and Manhart \cite{Schwertfirm_Manhart:2007} for finite $Re_\tau\geqslant180$, $Pe_\tau=Sc\,Re_\tau\geqslant20$, and $Sc\leqslant100$.
This relation is based on boundary-layer theory and is, thus, superior to any purely empirical relation (like the one from Sleicher and Rouse \cite{Sleicher_Rouse:1975}).
For small asymptotic $Sc<20\,Re_\tau^{-1}$ ($Pe_\tau<20$), there is currently no theory available but ODT consistently extrapolates $Sh$ to the diffusive limit, $Sh\to2$. 
For high asymptotic $Sc$, ODT results fall between the Dittus--Boelter, $Sh\sim Sc^{2/5}$, and Colburn, $Sh\sim Sc^{1/3}$, scalings but they are closer to the former than the latter, albeit latter is approached by the relation form Schwertfirm and Manhart.
Investigating $Sh(Pe)$ and $Sh(Re)$ for $Sc=0.025$ and $0.71$, respectively, revealed that the model obeys the same asymptotic scaling relations as available reference data \cite{Abe_Antonia:2019} for high $Pe>1000$.
ODT has limited predictive capabilities for $Pe<1000$ but stays within physical bounds.
Complex three-dimensional flow structures are not fully captured by the model but it can now be more confidently applied to multiphysical wall-bounded turbulent flows throughout the relevant parameter space.

\section*{Acknowledgements}
 
We thank Alan R. Kerstein for fruitful discussion.

\section*{Funding}

M.K. and H.S. acknowledge financial supported by the European Regional Development Fund (EFRE), grant number StaF~23035000.

\section*{Code and data availability}

The fully-adaptive ODT implementation used for this study is publicly available at the following URL: \url{https://github.com/BYUignite/ODT}\;; data can be made available upon request.

\appendix

\section{\label{sec:odt-appendix} Stochastic turbulence modeling by ODT}

In the following, we give a brief but complete description of the lower-order stochastic model to aid the discussion of the results for the passive scalar.
The starting point is given by the reduced-dimensional stochastic conservation equations~(\ref{eq:gov}\textit{a,b}), which are repeated here for convenience,
\begin{subequations}
 \renewcommand{\theequation}{\theparentequation \textit{\alph{equation}}}
 \label{eq:gov-appendix}
 \begin{align}
  \frac{\partial \boldsymbol{u}}{\partial t} + \sum_{t_\text{e}} \mathcal{E}_{\boldsymbol{u}}(\boldsymbol{u})\,\tilde{\delta}(t-t_\text{e}) &= \nu \frac{\partial^2 \boldsymbol{u}}{\partial y^2} - \frac{1}{\rho} \frac{\text{d}\bar{p}}{\text{d}x}\,\boldsymbol{e}_x \;,
  \\
  \frac{\partial \theta}{\partial t} + \sum_{t_\text{e}} \mathcal{E}_{\theta}(\boldsymbol{u})\,\tilde{\delta}(t-t_\text{e}) &= \Gamma \frac{\partial^2 \theta}{\partial y^2} + s_\theta \;.
 \end{align}
\end{subequations}
We adopt the dimensional formulation in order to highlight the physical aspects of the stochastic model formulation.
The deterministic processes evolve continuously such that they are represented by parabolic partial differential equations in between any two eddy events.
The numerical treatment is straightforward and not detailed here.
We only mention that a mesh-adaptive finite-volume method is used \cite{Lignell_etal:2013}.
Below, we elaborate the formulation of the stochastic terms and the underlying map-based advection modeling strategy that utilizes discrete eddy events.

\subsection{\label{sec:eddy} Map-based advection modeling by discrete eddy events}

Discrete eddy events are used to formulate the stochastic terms in equations~(\ref{eq:gov-appendix}\textit{a,b}).
This involves two mathematical operations to represent the effects of turbulent advection and pressure fluctuations.
When an eddy event is selected, the variables at location $f(y)$ are instantaneously replaced by the values at mapped location $y$.
For the scalar and the velocity vector, these operations~\cite{Kerstein_etal:2001} are given by
\begin{subequations}
  \renewcommand{\theequation}{\theparentequation \textit{\alph{equation}}}
  \label{eq:eddy2}
  \begin{align}
    \mathcal{E}_{\theta}:~ \theta(y) \to \theta''(y) &= \theta\big(f(y)\big)\;,
    \\
    \mathcal{E}_{\boldsymbol{u}}:~ \boldsymbol{u}(y) \to \boldsymbol{u}''(y) &= \boldsymbol{u}\big(f(y)\big) + \boldsymbol{c}(\boldsymbol{u};\alpha)\, K(y) \;,
  \end{align}
\end{subequations}
where $f(y)$ denotes a mapping function, $K(y)=y-f(y)$ a kernel function, and $\boldsymbol{c}=(c_1,c_2,c_3)^\text{T}$ the coefficients that are used to model the effect of pressure-velocity couplings based on the momentary velocity vector $\boldsymbol{u}(y)$ and the ODT model parameter $\alpha$.
This parameter, or the coefficients $c_i$, respectively, control the redistribution of the kinetic energy among the velocity components $u_i$, where $i\in\{1,2,3\}$.

In Navier--Stokes turbulence, the turnover of a single eddy locally increases the gradients of the flow variables on the length scale of that eddy.
This process is addressed in ODT by the triplet map~\cite{Kerstein:1999}.
For an eddy event of size $l$ occurring at lower eddy-edge location $y_0$, the triplet map affects the interval $[y_0,y_0+l]$ and is given by
\begin{equation}
  f(y) = y_0 + \left\{
  \begin{array}{ll}
         3(y-y_0) &\ \text{for}\quad y-y_0 \in \left[    0,  l/3 \right], \\%
    2l - 3(y-y_0) &\ \text{for}\quad y-y_0 \in \left[  l/3, 2l/3 \right], \\%
    3(y-y_0) - 2l &\ \text{for}\quad y-y_0 \in \left[ 2l/3,  l   \right], \\%
      y-y_0       &\ \text{otherwise.}
  \end{array}
  \right.
  \label{eq:triplet}
\end{equation}

In the dynamically adaptive implementation of the ODT model that is used here, profiles of the flow variables are spatially compressed by a factor of three, then the copies are sequentially arranged to fill the eddy size interval with the central copy spatially reversed in order to ensure continuity~\cite{Lignell_etal:2013}.
Irrespective of the implementation details, the important properties of the triplet map are that it is (i)~measure-preserving, and (ii)~does not introduce discontinuities along the ODT line.
These two aspects are important for the conservation properties of the method.
Note that, due to the triplet map, kinetic energy is brought from large to small scales in a scale-local fashion, which is consistent with the behavior of the direct energy cascade that is exhibited by three-dimensional turbulence.

Next, we continue with modeling of pressure-velocity couplings in the momentum equation.
The last term in equation~(\ref{eq:eddy2}\textit{b}) models the effect of a fluctuating pressure gradient.
The kernel function $K(y)$ is a measure for the map-induced fluid displacement, which means that the kernel function is a model for the microstructure of turbulent eddies, and the coefficient vector components $c_i$ scale the efficiency of the inter-component kinetic energy transfer.

The change of the kinetic energy in the $i^{\text{th}}$ velocity component due to the application of an eddy event, as expressed by equation~(\ref{eq:eddy2}b), is given by
\begin{equation}
 \Delta E_i = \dfrac{1}{2l} \int_{y_0}^{y_0+l} 
 \left[ 
   u_i^{\prime\prime 2}(y)
   - u_i^2(y) 
 \right] \,\text{d}y\;.
 \label{eq:dE}
\end{equation}
Energy conservation requires that the sum of the changes vanishes: $\Delta E_1 + \Delta E_2 + \Delta E_3 = 0$~\cite{Kerstein_etal:2001}.
The $c_i$ are obtained by a maximization of the inter-component kinetic energy transfer (that is, $-\Delta E_i$) with respect to $c_i$.
This yields
\begin{equation}
   c_i = \frac{1}{K_K}
     \left[ - u_{K,i} + \text{sgn}(u_{K,i}) \sqrt{ (1-\alpha) u_{K,i}^2 + \frac{\alpha}{2} \left( u_{K,j}^2 + u_{K,k}^2\right) } \,\right]\;,
  \label{eq:ci}
\end{equation}
where $u_{K,i}=\int u_i\big(f(y)\big)\, K(y) \,\text{d}y$ denotes the kernel-weighted components of the velocity vector, $K_K=\int K^2(y) \,\text{d}y$ the squared kernel, which is related to the map-induced fluid displacement, the indexes $(ijk)$ are permutations of $(123)$, and $\alpha$ is a model parameter that controls the efficiency of the inter-component energy transfer due to fluctuating pressure gradient forces. 
The parameter $\alpha$ specifies the fraction of the available (extractable) kinetic energy that is actually used for redistribution~\cite{Kerstein_etal:2001}.
It takes values between $0$ (no redistribution) and $1$ (maximal redistribution).
$\alpha=2/3$ corresponds to an equal partitioning of energy among velocity components, that is, a relaxation to locally homogeneous isotropic turbulence~\cite{Kerstein_etal:2001}, and is commonly applied in ODT simulations. 
Here we select $\alpha=1/6$ from a calibration of near-surface velocity fluctuations that are addressed in sections~\ref{sec:val} and~\ref{sec:jpdf}, as well as \ref{sec:cal}.

\subsection{\label{sec:selection} Stochastic selection of eddy events}

Eddy events are characterized by three random variables: the eddy size $l$, the lower eddy-edge location $y_0$, and the time $t$ of occurrence.
In theory, these variables can be sampled from an eddy-rate distribution $\lambda$, whereby $\lambda(l,y_0;t)\,\text{d}l\,\text{d}y_0\,\text{d}t$ gives the number of eddy events in the size range $[l, l+\text{d}l]$, the position range $[y_0, y_0 + \text{d}y_0]$ and the time interval $[t,t+\text{d}t]$.
This distribution is expensive to compute, and depends on the evolving flow state, further complicating its construction and sampling.

In practice, eddies are sampled from $\lambda(l,y_0;t)$ using the efficient thinning-and-rejection method~\cite{Lewis_Shedler:1979,Papoulis_Pillai:2002}.
In this method, temporal stochastic sampling exploits the assumption of independent (Markovian $\tilde{\delta}$-correlated) eddy events by scheduling the times $\left\{t_k\right\}_{k=1}^{N_\text{e}}$ with the aid of a marked Poisson process.
At this point, the stochastic dynamics would be fully random, which is in contrast to chaotic but deterministic turbulence dynamics. 
Therefore, non-Markovian memory is introduced into the sampling by accounting for the eddy-available energy based on the momentary flow state.
This energy has been worked out above in \ref{sec:eddy}. 
The relevant quantity that connects stochastic dynamics and flow physics is the eddy turnover time $\tau$, which is given in equation~(\ref{eq:eddyTau}).  

The eddy turnover time can be evaluated for any candidate eddy event, which implies a distribution function $\tau(l,y_0;t)$ that parametrically depends on time $t$ due to the eddy energetics for the momentary flow state. 
It is, hence, favorable to express the eddy-rate distribution $\lambda(l,y_0;t)$ in terms of the eddy size, $l$, and the eddy turnover time, $\tau$, as \cite{Kerstein:1999}
\begin{equation}
 \lambda(l,y_0;t)=\dfrac{C}{l^2\,\tau(l,y_0;t)} \;.
 \label{eq:eddyRate}
\end{equation} 
where $C$ is a proportionality constant (model parameter) that scales the eddy rate given the eddy turnover time for the selected eddy size and location. 
The momentary rate of all eddies, $\Lambda$, is given by the sum over all possible eddy sizes and locations as
\begin{equation}
  \Lambda(t) = \int_{L} \int_{L} \lambda(l,y_0;t)\,\text{d}l\,\text{d}y_0 \;.
 \label{eq:totalEddyRate}
\end{equation}
Eddy events can be as large as the ODT domain size $L$ ($L=2\delta$ for channel flows) and may, in general, occur anywhere along it.
Additional physical considerations (see below) may further constrain the eddy size. 
The momentary joint probability density function (jPDF) of $l$ and $y_0$ is given by $p(l,y_0;t) =
\lambda(l,y_0;t)/\Lambda(t)$.
Subsequent eddy-event occurrences are deemed independent from each other such that they can be sampled from a discrete Poisson process in time with momentary rate $\Lambda$ corresponding to a given (`frozen') flow state. 
In principle, eddy-event occurrences can be sampled based on Poisson statistics, and the values $l$ and $y_0$ can be sampled from the jPDF $p(l,y_0;t)$ at a selected time $t$.
In practice, this would be too costly since $p(l,y_0;t)$ is constantly changing in an unpredictable way as the flow evolves.
Therefore, for any scheduled eddy event occurrence $t=t_\text{e}$, the jPDF $p(l,y_0;t)$ is approximately factored into assumed candidate distribution functions,
\begin{equation}
 p(l,y_0;t_\text{e}) \approx g_1(l)\,g_2(y_0) \;,
 \label{eq:eddyFactored}
\end{equation}
each of which depends only on a single variable.
Here, $g_1(l)$ is a skewed unimodal distribution with a peak at around the viscous length scale, that is, $l_\text{p}\approx\delta_\nu$.
This is motivated by the fact that very small eddy events will be subject to viscous suppression, whereas very large eddy events are rare.
The assumed PDF $g_1(l)$ was originally parameterized by Scott Wunsch as described in \cite{McDermott:2005} and is given by 
\begin{equation}
 g_1(l) = \dfrac{ -2l_\text{p} }{ l^2 }
        \left[ \dfrac{ \exp(-2l_\text{p}/l) }
                     { \exp(-2l_\text{p}/l_\text{max}) - \exp(-2l_\text{p}/l_\text{min}) }
        \right] \;,
 \label{eq:gL}
\end{equation}
where $l_\text{min}$, $l_\text{max}$, and $l_\text{p}$ are user-specified minimum, maximum, and most probable eddy-event sizes, respectively, that can be adjusted for efficiency.
Since eddy events can in principle happen anywhere, the assumed PDF $g_2(y_0)=(L-l)^{-1}$ is simply uniform for the accessible lower-edge locations $y_0\in[0,L-l]$ of a selected size-$l$ eddy event.
It is worth to point out that the exact forms of the selected candidate distribution functions $g_1(l)$ and $g_2(y_0)$ only affect model efficiency but \emph{not} the results as long as the distribution functions cover the physically relevant parameter space of the turbulent flow (see~\cite{Kerstein:1999} for further discussion).
The physically relevant jPDF $p(l,y_0;t_\text{e})$ of implemented eddy events is approximately obtained with the aid of energetically constrained rejection sampling that is described next. 

We continue by writing the eddy turnover time in terms of the total extractable (shear-available) kinetic energy per unit mass, which, for the instantaneous velocity components $u_i(y,t)$ and a selected eddy event of size $l$, is given by
\begin{equation}
  \frac{l^2}{\tau^2} \sim \frac{1}{l^4} \sum_{i=1}^{3} u_{K,i}^2 - Z \frac{\nu^2}{l^2} \;.
  \label{eq:eddyEnergy}
\end{equation}
It follows from the construction that the kernel-weighted velocities can be summed instead of the kinetic energies.
This shows that the total extractable kinetic energy does not depend on the inter-component energy transfer, or the model parameter $\alpha$.
The last term in equation~\eqref{eq:eddyEnergy} is included to model the damping effects of the viscosity.
The model parameter $Z>0$ is used to suppress unphysically small eddy events through an energetic penalty condition~\cite{Kerstein:1999}.
Values $Z\gg1$ have been suggested for wall-bounded flows~\cite{Kerstein:1999,Schmidt_etal:2003} to make the model aware of the wall-attached viscous sublayer by effectively suppressing near-wall eddies below the critical local Reynolds number $Re_\text{c}=u_\text{e}l/\nu=\sqrt{Z}$, where $u_\text{e}=(u_K^2 + v_K^2 + w_K^2)^{1/2}$.

Finally, the eddy turnover time, $\tau$, can be computed under all circumstances from the instantaneous velocity vector components $u_i(y,t)$ once the location $y_0$ and size $l$ have been selected, that is,
\begin{equation}
  \frac{1}{\tau} = \sqrt{ \frac{1}{l^6} \left(u_K^2 + v_K^2 + w_K^2\right) - Z \frac{\nu^2}{l^4}} \;.
  \label{eq:eddyTau}
\end{equation}
This time scale is compared with the mean sampling time scale, $\tau_\text{s}$, to obtain the acceptance probability as
\begin{equation}
 p_\text{a} = \dfrac{\tau_\text{s}}{\tau\,l^2\,g_1(l)\,g_2(y_0)} < 1 
 \label{eq:pa}
\end{equation}
for any physically plausible eddy event.
Candidate eddy-event occurrences are sampled based on Poisson statistics by utilizing a marked Poisson process with rate $\tau_\text{s}^{-1}$.
The sampling time increment, $\tau_\text{s}$, is chosen so that the sampling rate exceeds $\Lambda$. 
For each of these eddy-event occurrences, $l$ and $y_0$ are sampled from $g_1(l)$ and $g_2(y_0)$, respectively, and a candidate eddy event is finally accepted with probability $p_\text{a} < 1$ due to oversampling.

It is sometimes important to suppress unphysically large eddy events, which may occur rarely in the sampling procedure but have significant effect on the mixing process as explained above.
A simple suppression mechanism is often sufficient for confined flows.
For fully-developed channel flow, turbulence statistics must be symmetric about the channel center.
Based on Prandtl's mixing length theory \cite{Prandtl:1925} and the attached-eddy model \cite{Townsend:1976,Marusic_Monty:2019}, it has been suggested \cite{Schmidt_etal:2003} to constrain the maximum eddy size by the channel half-height, that is, $l_\text{max}=\delta$, which is done here. 

\subsection{\label{sec:num} Remarks on the numerical solver implementation}

The stochastic terms denoted by $\mathcal{E}$ above in equations~(\ref{eq:gov-appendix}\textit{a,b}) are zero in between any two instantaneous eddy events so that a set of one-dimensional diffusion equations is solved continuously in time until the next eddy event occurs.
The time advancement is done here with the explicit Euler method.
The implicit Euler method has been used for a couple of cases, though with little effect on the simulation efficiency that is limited by the eddy sampling procedure.  
Higher order methods are implemented, but the instantaneous implementation of the stochastic mapping events nevertheless results in a first-order global scheme. 

Spatial discretization along the ODT domain is done with a Lagrangian finite-volume method on an adaptive grid~\cite{Lignell_etal:2013}.
The velocity vector and the passive scalar are located at the cell centers. 
The minimum allowed cell size, $\Delta y_\text{min}$, which is needed for dynamic re-meshing, is of the order of the Batchelor scale that can be estimated as $\eta_\text{B}/\delta\simeq Sc^{-1/2} Re^{-3/4}$, where $Re$ denotes the bulk Reynolds number, which can be estimated for fully-developed turbulent channel flow by inverting $Re_\tau\simeq0.18\,Re^{0.88}$ \cite{Pope:2000}.
The maximum allowed cell size is set to $\Delta y_\text{max}\simeq 20\,\Delta y_\text{min}$ in order to minimize any numerical transport.

In the adaptive mesh ODT formulation~\cite{Lignell_etal:2013}, the grid density increases in eddy regions during eddy events, consistent with the factor-of-three spatial compression of the triplet maps. This eddy region undergoes mesh adaption again at the end of the diffusive advancement process, and the local (versus domain-global) adaption is done to minimize numerical effects of the meshing operations. However, a given region is subject to mesh adaption if it has not had an eddy event and been adapted within some multiple, $\beta_\text{DA}$, of the diffusion time scale, where $\beta_\text{DA}=10$ is used.

\section{\label{sec:bc} Scalar forcing by isothermal and isoflux boundary conditions}

Two scalar forcing methods are considered in this study that are sketched in figure~\ref{fig:config}.
In the first, the scalar is prescribed by fixed wall values, $\theta(y=0)=\theta_\text{bot}$ at the bottom and $\theta(y=2\delta)=\theta_\text{top}$ at the top, respectively.
This case is denoted as CSV (constant scalar value).
The scalar forcing term vanishes for this configuration so that $s_\theta=0$ in equation~(\ref{eq:gov}\textit{b}).
The ODT domain is fixed at some location of the channel and flow statistics are gathered in time for an $M\geqslant1$ ensemble of independent flow realizations.

In the second forcing method, the scalar is prescribed by an isoflux wall-boundary condition and is denoted as CSF (constant scalar flux).
This configuration yields a linear increase of the bulk scalar property downstream,
\begin{equation}
  \Theta_\text{b} = q_\text{w} \, t = q_\text{w} \, x/U_\text{d} \;,
  \label{eq:tbulk}
\end{equation}
where $q_\text{w}$ is the (constant) prescribed surface scalar flux into the flow domain, $t$ the elapsed simulation time, $x$ the downstream distance, and $U_\text{d}$ a constant displacement velocity of the ODT domain.
With these boundary conditions a temporal ODT simulation would result in a solution that implies a displacement of the ODT domain downstream with constant mean velocity $U_\text{d}$, as indicated by two small horizontal arrows in figure~\ref{fig:config}(b).
This corresponds with previous applications of the \emph{temporal} ODT formulation (e.g.~\cite{Echekki_etal:2001,Klein_etal:2019}) where $U_\text{d}$ was prescribed by the bulk velocity.
In fact, this is also a common approach when post-processing temporal ODT simulation results for comparison to spatial experimental data.
We emphasize that the approach differs somewhat from the \emph{spatial} ODT formulation (e.g.~\cite{Kerstein:1999,Monson_etal:2016}) that uses the local streamwise velocity in spatially advancing the ODT line in an assumed steady flow that is, nevertheless, punctuated by instantaneous eddy events. 
However, the spatial ODT formulation can not be straightforwardly applied to internal flows \cite{Medina_etal:2019}. 

As a consequence of these limitations, we apply only the temporal formulation for this study. 
To facilitate this, the scalar transport equation is modified to compensate for the linear dependence in equation~\eqref{eq:tbulk} by a transformation of variables, as has been previously applied in DNS studies \cite{Kasagi:1992,Abe_Kawamura_Matsuo:2004}.

For the sake of argument, consider the scalar transport equation specialized to the temperature $T$,
\begin{equation}
  \dfrac{\partial T}{\partial t} + \boldsymbol{u} \boldsymbol{\cdot} \boldsymbol{\nabla} T = \Gamma\nabla^2 T \;,
  \label{eq:Ttran}
\end{equation}
where $\Gamma$ is the thermal (scalar) diffusivity. 
In statistically stationary channel flow with a constant and uniform wall heat flux applied, the mean temperature profile increases linearly with downstream distance $x$. 
We define $\theta=\bar{T}_\text{w}-T$, where $\bar{T}_\text{w}$ is the mean wall temperature, which increases linearly with $x$. 
This gives $T=\beta x + \bar{T}_{\text{w},0} - \theta$, where $\beta=\text{d}\bar{T}_\text{w}/\text{d}x$, which is constant. 
This relation is inserted into the transport equation~(\ref{eq:Ttran}) to yield
\begin{equation}
  \dfrac{\partial \theta}{\partial t} + \boldsymbol{u} \boldsymbol{\cdot} \boldsymbol{\nabla} \theta = \Gamma\nabla^2\theta + \beta u \;,
  \label{eq:TtranTheta}
\end{equation}
where the last term follows from $(\boldsymbol{u} \boldsymbol{\cdot} \boldsymbol{\nabla}) \beta x = \beta u$. 
For application in ODT, equation~(\ref{eq:gov}) is solved rather than equation~(\ref{eq:TtranTheta}), with source term $s_\theta=\beta u$ and the stochastic term $\sum_{t_\text{e}}\mathcal{E}_\theta(\boldsymbol{u})\,\tilde{\delta}(t-t_\text{e})$ formally replacing the advection term $\boldsymbol{u} \boldsymbol{\cdot} \boldsymbol{\nabla} \theta$.

Note that $\beta$ can be chosen arbitrarily. 
Any constant multiplier of $\beta$ will not affect statistics of $\theta^+$. 
The friction temperature is given by $T_\tau=\bar{q}_\text{w}/(\rho c_p u_\tau)$, which is consistent with equation~(\ref{eq:nondim}b) for $T=\theta$. But the mean heat (scalar) flux $\bar{q}_\text{w}$ is proportional to $\beta$, as a simple energy (scalar mass) balance over the channel shows. 
Hence, inner scaling of equation~(\ref{eq:TtranTheta}) with $T_\tau$ to give a transport equation for $\theta^+$ results in a cancellation of $\beta$ in the source term.
As a consequence, Dirichlet boundary conditions for $\theta$ are used with $\theta(y=0)=\theta(y=2\delta)=0$.

This formulation of the isoflux boundary conditions is particularly convenient since a standard transport equation can be solved with a simple source term. This also allows application of the temporal ODT formulation without requiring advection of the ODT line, as noted above. 

In equation~(\ref{eq:Ttran}), $u$ in the source term is taken as the local and momentary ODT streamwise velocity, $u(y,t)$. 
Results of this configuration are presented in sections~\ref{sec:tmean} and \ref{sec:jpdf}.
The majority of the results shown use fixed-value (CSV) wall-boundary conditions and only a few are based on the isoflux (CSF) boundary condition for the purpose of one-to-one comparison with reference DNS where necessary.
We justify this, as discussed throughout section~\ref{sec:results}, by boundary layer similarity that is primarily relevant for the modeling of fluctuating surface fluxes at high Reynolds and Peclet numbers.

\section{\label{sec:cross-corr} Computation of model-resolved turbulent fluxes}

The wall-normal components of the Reynolds stress tensor, $\overline{u_i'u_2'}\equiv\overline{u_i'v'}$, and the turbulent scalar flux, $\overline{\theta'v'}$, are implicitly resolved along the ODT domain.
In order to compute these cross-correlation terms within the model we have to take into account that eddy events have taken the role of turbulent advection.
Fluid is displaced along the ODT domain by application of the triplet map given in equation~\eqref{eq:triplet}. 
Turbulent advective transport is thus an ensemble effect of triplet map applications across a range of scales.
These mappings are instantaneous but well defined in location, $y_0$, and size, $l$.

A measure for the duration of a physical eddy turnover to yield a similar manipulation of property profiles is, thus, given by the eddy turnover time, $\tau$, from equation~\eqref{eq:eddyTau}.
This time scale is readily available from the sampling procedure and we estimate the advecting fluctuation velocity as $l/\tau$.
Now we have to separate the effects of triplet map applications from other processes that change property profile.
This is achieved by a conditional averaging procedure which, however, can \emph{not} be done alone with an instantaneous property profiles.
Instead, a finite time interval (size $\Delta t_\text{e}$) is used to perform conditional eddy statistics in order to estimate the model-resolved turbulent fluxes, that is, cross-correlations that involve $v'$.

For the statistically stationary state, the ensemble effect of stochastic eddy events may be written as
\begin{equation}
  \frac{\partial \overline{v'\phi'}}{\partial y}  \approx -\frac{\Delta \phi_\text{e}}{\Delta t_\text{e}}\;,
 \label{eq:reystress-div}
\end{equation}
where $v'$ denotes the `real' velocity fluctuation in the direction of the ODT line and, correspondingly, $\phi'$ the `real' fluctuation of a transported flow variable, that is, either a velocity component or the scalar.
The cumulative map-induced (`turbulent') change, $\Delta \phi_\text{e}$, of variable $\phi$ is thus given by
\begin{equation}
 \Delta \phi_\text{e}(y) = \sum_{k=1}^{N_\text{e}} \big[ \phi''(y,t_k) - \phi(y,t_k) \big] \frac{t_k-t_{k-1}}{\Delta t_\text{e}} \;,
 \label{eq:reystress-du}
\end{equation}
where index $k$ counts $N_\text{e}$ eddy events during $\Delta t_\text{e}=t_{N_\text{e}}-t_0$, in which $t_0$ is the start and $t_{N_\text{e}}$ the end time of the statistics gathering.
Furthermore, $\phi''$ represents the map-induced changes of $\theta''$ or $\boldsymbol{u}''$ as defined by equations~(\ref{eq:eddy2}\textit{a,b}).

At last, equation~\eqref{eq:reystress-div} is integrated along the ODT domain to yield the turbulent flux of the transported variable $\phi$.
We obtain
\begin{equation}
 \overline{ v'\phi' }(y) \simeq
   - \int_0^y
     \left\langle \frac{\Delta \phi_\text{e}}{\Delta t_\text{e}}(y') \right\rangle_M \,\text{d}y' \;
 \label{eq:reystress}
\end{equation}
as wall-normal profile.
Here, we have introduced the ensemble average that is denoted by $\langle\cdot\rangle_M$ and encompasses $M=1,2,\ldots$ flow realizations.
Ensemble averaging is optional but may be used to converge the fluctuation statistics faster by running several ODT simulations of one case in parallel.

\section{\label{sec:cal} Model calibration for the velocity boundary layer}

We now aim to evaluate the general model fidelity in terms of low-order velocity statistics in the turbulent boundary layer for the calibrated model parameters by comparison with available reference data.

\begin{figure}[tp]
  \centering
  \includegraphics[height=52mm]{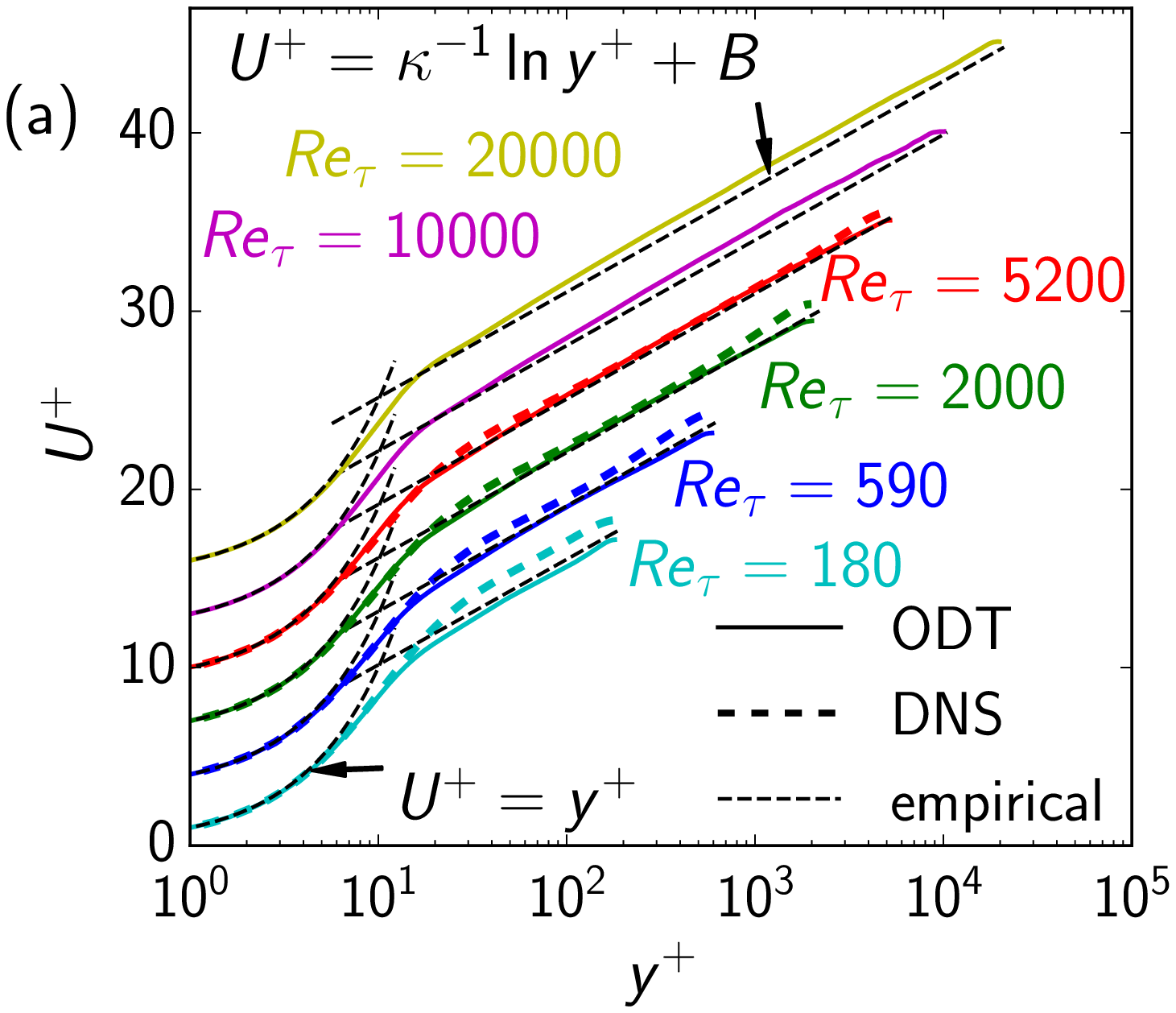} 
  \includegraphics[height=52mm]{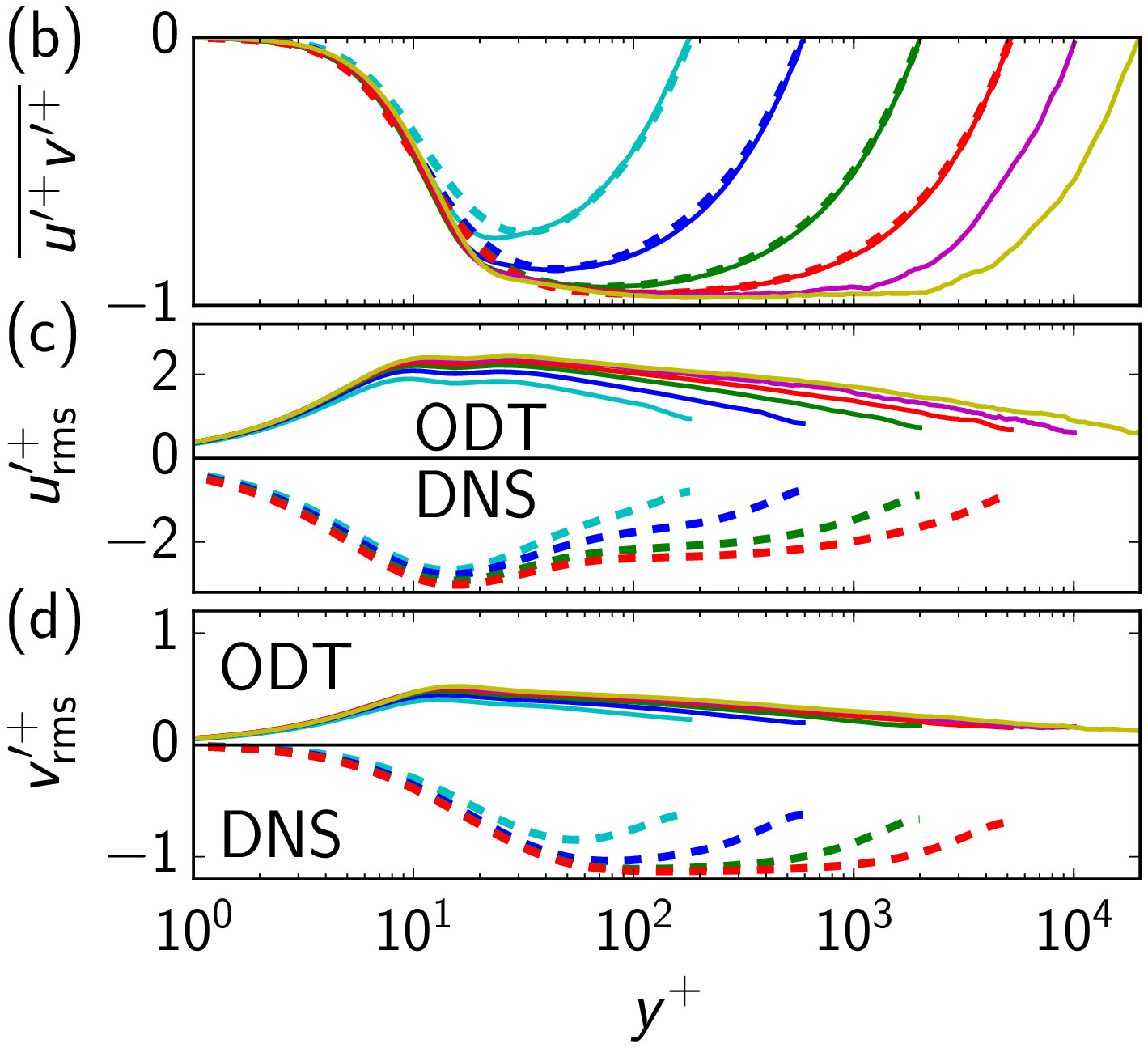}
  \caption{%
    (a)~Mean stream-wise velocity, $U^+$ (shifted by $\mathrm{\Delta} U^+=3$ for each increment in $Re_\tau$);
    (b)~specific wall-normal turbulent momentum flux, $\overline{u^{\prime +}v^{\prime +}}$;
    (c)~stream-wise, $u^{\prime +}_\text{rms}$; and 
    (d)~wall-normal, $v^{\prime +}_\text{rms}$, root-mean-square (rms) fluctuation velocity components for various $Re_\tau$ that are distinguished by colors. 
    Panels~(c) and (d) are vertical `back-to-back' plots with reference data (negative values, dashed) mirrored at the $y^+$ axis for better visibility. 
    Reference DNS results are from \cite{Lee_Moser:2015,Moser_etal:1999}.
    The empirical law-of-the-wall is given for high asymptotic $Re_\tau$ yielding the log-layer parameterization coefficients $\kappa=0.389$ and $B=4.23$ \cite{Marusic_etal:2010}.
  }
  \label{fig:vel}
\end{figure}

Figure~\ref{fig:vel}(a) shows normalized mean velocity profiles, $U^+=U/u_\tau$, for various $Re_\tau$ for the calibrated model.
Simulated mean velocity profiles are consistent with the empirical law-of-the-wall of high asymptotic $Re_\tau$ \cite{Marusic_etal:2010}.
This demonstrates the model consistency and convergence for highly turbulent flows with large Reynolds number. 
The ODT model accurately captures not only the log layer, but also the linear (viscous) layer right at the wall.
However, it does not fully capture the buffer and outer layers of the reference DNS \cite{Lee_Moser:2015,Moser_etal:1999}.

Figure~\ref{fig:vel}(b) shows the normalized Reynolds stress component $\overline{u^{\prime +} v^{\prime +}}$, which is in very good agreement with the reference DNS \cite{Lee_Moser:2015,Moser_etal:1999}.
This implies model consistency in the log and linear layers with the mean velocity and mean shear including wall-normal stresses.

Figures~\ref{fig:vel}(c) and \ref{fig:vel}(d) show the normalized root-mean-square (rms) fluctuation velocity components, $u_{\text{rms}}^{\prime +}=(\overline{u^2} - U^2)^{1/2} /u_\tau$ and $v_{\text{rms}}^{\prime +}=(\overline{v^2})^{1/2} /u_\tau$, respectively, as vertical `back-to-back' plots with reference DNS data \cite{Lee_Moser:2015} mirrored at the horizontal axis (negative fluctuation velocity) for visibility.  
In ODT, the fluctuation velocities are generally underestimated, which is a known modeling artifact (e.g.~\cite{Kerstein:1999,Kerstein_etal:2001,Rakhi_etal:2019,Schmidt_etal:2003}).
The location of the inner (near-wall) peak of $u^{\prime +}_\text{rms}$ is captured but the peak itself is degraded due to the micro-structure modeling in the discrete eddy events \cite{Lignell_etal:2013}.
The outer peak in $u^{\prime +}_\text{rms}$ that emerges at high $Re_\tau$ is only marginally captured by the model \cite{Rakhi_etal:2019}.

The wall-normal fluctuation variance, $v^{\prime +}_\text{rms}$, which is shown in figure~\ref{fig:vel}(d), is generally less well captured than $u^{\prime +}_\text{rms}$, except for the surface layer region with $y^+<5$ for the model setup used.
The latter aspect is crucial with respect to capturing transient surface fluxes which is the primary motivation for this study that focuses on wall-normal transport and models stream- and span-wise contributions in the same way by a simple kernel mechanism that is given in equation~(\ref{eq:eddy2}b).
Note that eddy turbulent advection in ODT is performed by the triplet map \eqref{eq:triplet} such that $v'$ is \emph{not} a turbulent advecting velocity.
Instead, the fluctuation variance $\overline{v^{\prime 2}}$ (and, hence, also $v^{\prime +}_\text{rms}$) is a measure for the contribution from the $v$ velocity component to the model-resolved eddy turbulent kinetic energy.
It is therefore less surprising that the turbulent flux $\overline{u'v'}$ is well captured by map-based advection modeling even though $u'$ and $v'$ (as well as $w'$) are not fully captured \cite{Schmidt_etal:2003}. 

Next, we analyze the turbulent kinetic energy (TKE) in order to quantify the model-resolved fluctuating momentum transport across the velocity boundary layer.
The TKE budget balance is given by (see e.g.~\cite{Pope:2000})
\refstepcounter{equation}
$$
  \frac{\partial k}{\partial t}
   + \boldsymbol{U} \boldsymbol{\cdot} \boldsymbol{\nabla} k
   = P + \varepsilon + D + T \;,
  \eqno{(\theequation)}
  \label{eq:tke}
$$
where $k=\overline{u_i'u_i'}/2$ denotes the TKE and $P$ the turbulent production, $\varepsilon$ the dissipation, $D$ the diffusive transport, and $T$ the turbulent transport of the TKE.
The left-hand side vanishes for statistically stationary channel flow, and the terms on the right-hand side only retain their wall-normal contributions within the ODT modeling framework.
Following \cite{Kerstein:1999}, we obtain the following model-resolved terms:
\begin{subequations}
 \renewcommand{\theequation}{\theparentequation \textit{\alph{equation}}}
 \label{eq:tke-odt}
 \begin{align}
  P &= -\overline{v'u'} \,\frac{\text{d} U}{\text{d} y} \;,
  \\
  \varepsilon &= -\nu \overline{ \frac{\partial u_i'}{\partial y}   \frac{\partial u_i'}{\partial y} } \;,
  \\
  D &= \nu \frac{\text{d}^2 k}{\text{d} y^2} \;,
  \\
  T &= -\frac{\text{d} \,\overline{v'u_i'u_i'}}{\text{d} y} \;.
 \end{align}
\end{subequations}
The corresponding dimensionless expressions for these terms are obtained by division with $u_\tau^4/\nu$, which is indicated by the super-script `$+$'.
Note that the fluctuating pressure transport is not directly resolved by ODT.
However, pressure transport is implicitly modeled in the fluctuating terms that are stochastically modeled.
Note further that we compute cross-correlations, like $\overline{u'v'}$, directly but diagnostically from conditional eddy event statistics.
The procedure is described in \ref{sec:cross-corr}.

\begin{figure}[tp]
  \centering
  \includegraphics[height=52mm]{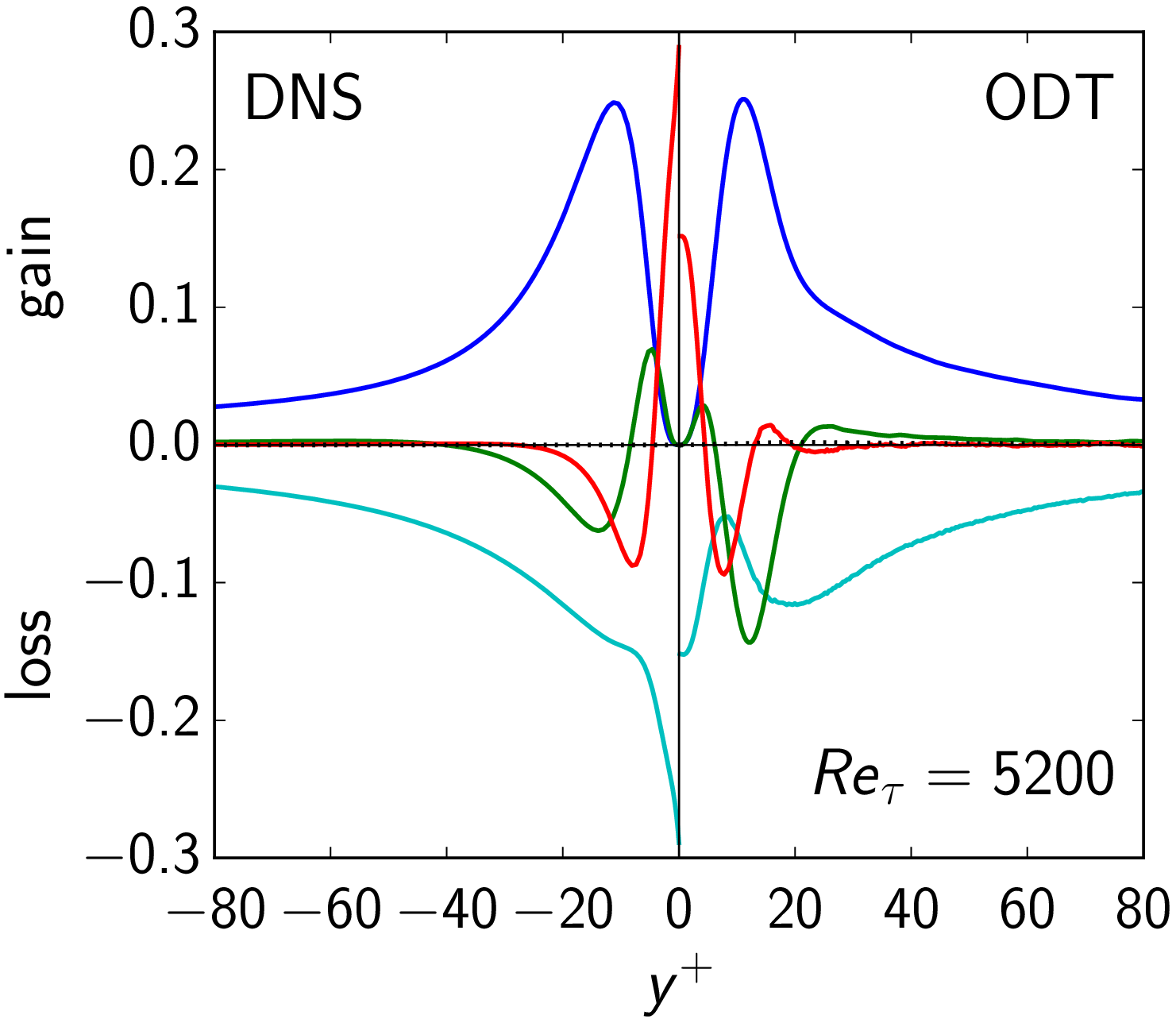} 
  \includegraphics[height=52mm]{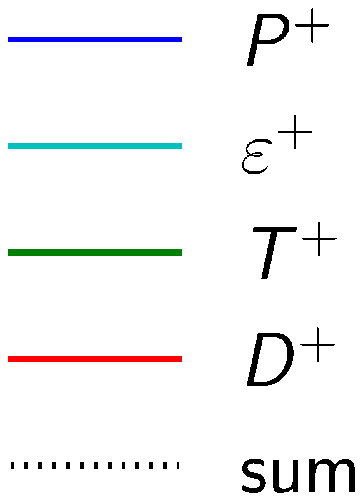}
  \caption{%
    `Back-to-back' plot of the turbulent kinetic energy (TKE) budget balance for $Re_\tau=5200$ showing ODT results for $y^+\geqslant0$ and corresponding DNS from \cite{Lee_Moser:2015} for $y^+\leqslant0$.
  }
  \label{fig:tke}
\end{figure}

Figure~\ref{fig:tke} shows the TKE budget balance for the model-resolved terms together with reference data.
One can see that all terms are qualitatively captured.
There is good quantitative agreement for the production, but the dissipation and transport terms agree quantitatively only towards the bulk.
Hence, the momentum transport across the buffer layer, $5<y^+<30$, is somewhat different in the lower-order stochastic representation of the turbulence dynamics.
Nevertheless, we emphasize that the model is consistent in exhibiting a balance of the resolved budget terms (dotted line).





\end{document}